\documentclass[letterpaper]{article}
\usepackage{aaai25}
\usepackage{times}
\usepackage{helvet}
\usepackage{courier}
\usepackage[hyphens]{url}
\usepackage{graphicx}
\usepackage{natbib}
\usepackage{caption}
\usepackage{amsfonts}
\usepackage{amsmath}
\usepackage{amssymb}
\usepackage[
  breaklinks=true,
  colorlinks=true,
  linkcolor=RedOrange,
  urlcolor=ForestGreen,
  citecolor=NavyBlue,
  anchorcolor=ForestGreen,
  backref=page,
]{hyperref}       
\usepackage{mathtools}
\usepackage{amsthm}
\usepackage{enumitem}
\usepackage{multirow}
\usepackage{mathtools}
\usepackage{tabularx}
\usepackage{pifont}
\usepackage{booktabs} 
\usepackage[dvipsnames,table]{xcolor} 
\usepackage[capitalize,noabbrev]{cleveref}

\setlist{leftmargin=*,itemsep=0pt,topsep=0pt}
\newcolumntype{C}{>{\centering\arraybackslash}X}
\newcolumntype{Y}{>{\raggedright\arraybackslash}X}

\newcommand{\demb}{\ensuremath d_{\mathrm{emb}}}
\newcommand{\emb}{\ensuremath x}
\newcommand{\nval}{\ensuremath n_{\mathrm{val}}}
\newcommand{\bR}{\ensuremath \mathbb{R}}
\newcommand{\xhdr}[1]{\paragraph{#1}}

\title{Whole Genome Transformer for Gene Interaction Effects\\ in Microbiome Habitat Specificity}
\author{
    Zhufeng Li\textsuperscript{\rm 1,2,3},
    Sandeep S Cranganore\textsuperscript{\rm 4,5},
    Nicholas Youngblut\textsuperscript{\rm 6},
    Niki Kilbertus\textsuperscript{\rm 1,2,3}
}
\affiliations{
    \textsuperscript{\rm 1}Technical University of Munich,\:
    \textsuperscript{\rm 2}Helmholtz Munich,\:
    \textsuperscript{\rm 3}Munich Center for Machine Learning (MCML),\:
    \textsuperscript{\rm 4}Forschungszentrum Jülich,\:
    \textsuperscript{\rm 5}Technical University of Vienna,\:
    \textsuperscript{\rm 6}Max Planck Institute for Biology, Tübingen\\
    \{zhufeng.li,niki.kilbertus\}@tum.de
}

\begin{document}

\maketitle

\begin{abstract}
Leveraging the vast genetic diversity within microbiomes offers unparalleled insights into complex phenotypes, yet the task of accurately predicting and understanding such traits from genomic data remains challenging. We propose a framework taking advantage of existing large models for gene vectorization to predict habitat specificity from entire microbial genome sequences. Based on our model, we develop attribution techniques to elucidate gene interaction effects that drive microbial adaptation to diverse environments. We train and validate our approach on a large dataset of high quality microbiome genomes from different habitats. We not only demonstrate solid predictive performance, but also how sequence-level information of entire genomes allows us to identify gene associations underlying complex phenotypes. Our attribution recovers known important interaction networks and proposes new candidates for experimental follow up.
\end{abstract}

\begin{links}
    \link{Code}{https://github.com/zhufengli/prokaformer}
\end{links}

\section{Introduction and related work}\label{sec:intro}

\xhdr{Machine learning (ML) on genetic data.}
Determining how gene-gene interactions influence certain traits, health, and disease has been a longstanding challenge for biologists and medical researchers \citep{Gilbert-Diamond2011-xh,WAN2010325}.
Modern high-throughput sequencing techniques such as massively parallel methods \citep{Ronaghi1996-qs,NYREN1993171,nayfach2021genomic} or single cell RNA sequencing \citep{hwang2018single,jovic2022single} together with recent developments in transformer-based models \citep{vaswani2017attention}, which nowadays operate on sequences lengths up to 100,000 \citep{Avsec2021-xz} or even 1 million \citep{nguyen2023hyenadna} base pairs, allow for modeling highly complex sequence diversity spanning large sections of the genome.

Within this paradigm, \citet{Jumper2021-zl} achieved state of the art in protein folding predictions, \citet{Avsec2021-xz} identified enhancer-promoter interactions with unprecedented accuracy, and \citet{li2023genomic,Avsec2021-xz} demonstrate promising results on gene regulatory network inference.
The potential impact on human health has also inspired large-scale concerted industry efforts into building large transformer models that can perform multiple relevant tasks at once. For instance, in a sequence of papers \citep{rives2019biological,rao2020transformer,rao2021msa,meier2021language,hsu2022learning,lin2022language,lin2023evolutionary}, a collection of models was released---dubbed Evolutionary Scale Modeling (ESM)---that perform tasks from protein design (beyond natural proteins) and (inverse) protein folding to variant-, function-, and property-prediction. \citet{consens2023transformers,choi2023transformer} provide detailed overviews of recent deep-learning (in particular transformer) based models for the genome and what they are capable of.

\xhdr{Importance of the microbiome.}
Bacteria and archaea are often heavily underrepresented in deep learning models trained on genetic data \citep{zhou2023dnabert,dalla2023nucleotide}. While modeling human genetic diversity has many direct implications for human health \citep{sapoval2022current,clapp2017gut}, developing models that incorporate the vast genetic diversity across the microbial tree of life may lead to similar benefits, such as the development of novel microbiome therapeutics, inferring the health benefits of microbe-produced metabolites, and predicting the evolution of antibiotic resistance \citep{hernandez2022machine}. Unlike the relatively static nature of the human genome, the microbiome is highly dynamic, adapting to environmental changes and interactions with its host or environment \citep{lloyd2017strains,ducarmon2023remodelling}. The plasticity of the microbiome could be harnessed to treat disease more easily via microbiome interventions versus gene- or immuno-therapy \citep{schupack2022promise,ratiner2023utilization}.

While works like ESM \citep{lin2023evolutionary} and LookingGlass \citep{hoarfrost2022deep} included a large degree of known microbial diversity, these models are limited to single genes or short DNA sequences of 100 to 200 base pairs \citep{hoarfrost2022deep}.
Moreover, microbial genes are often arranged in operons that are co-regulated and often form protein complexes \citep{cao2019door}. Modeling large segments of the genome can thus incorporate much more genotypic complexity than models trained on single genes or short DNA sequences \citep{wei2024annoview,nguyen2023hyenadna,cheifet2019genomics}.

Predicting phenotype from genotype is quite challenging in the context of the microbiome. First, the majority of microbial genome assemblies are not complete \citep{parks2022gtdb,chklovski2023checkm2}, and instead comprise 10's to 1000's of genome fragments (contigs). Even among individual genomes belonging to the same species, genomes can differ substantially in genomic content and arrangement \citep{rouli2015bacterial,lapierre2009estimating}; thus, the ordering of contigs usually cannot be inferred from closely related, completely assembled genomes. Second, microbial genome databases under-represent microbial diversity, especially microbes that are rare in well-studied environments or microbes only found in understudied environments \citep{brewster2019surveying,pavlopoulos2023unraveling}. Third, the cellular functioning of most microbial genes and non-coding elements is unknown, which has led to initiatives to uncover this ``microbial dark matter'' \citep{hoarfrost2022deep,pavlopoulos2023unraveling}; however, much work is still needed. This work is especially challenging, given that many microbes cannot be cultivated \citep{almeida2021unified}, and genetic tools only exist for a small subset of cultivatable microbes \citep{marsh2023toward}. Fourth, microbial phenotypes are often difficult to measure, given the challenge to isolate and measure the traits of individual strains. Complex phenotypes, such as microbial habitat, may involve a number of factors, including many cellular processes produced by a multitude of genes and regulatory elements.

\xhdr{Existing `genotype to phenotype' methods.} 
A number of approaches have been used to determine microbial phenotypes from genomic data.
The most prominent are homology-based methods in which the function of a gene (or other genetic element) is inferred by a sequence similarity search to references with characterized functions.
This approach is challenged by a lack of characterized references and the often incorrect assumption that sequence similarity predicts functional similarity.
A similar approach is genome-wide association (GWAS) of nucleotide-level variations among very closely related organisms to infer phenotype based on how genetic variation correlates to characterized phenotypic variation \citep{de2018complex,collins2018phylogenetic,lees2020improved,yang2023evolink}.
Such methods often require many closely related individuals (e.g., intra-species) with matched high quality genome assemblies and characterized phenotypes.
Another approach is the use of phylogenies to infer phenotypes of characterized sections of the evolutionary tree based on relatedness to characterized representatives.
This approach is challenged by the difficulties of inferring accurate phylogenies, obtaining adequate numbers of phenotypically characterized representatives, and assuming that evolutionary relatedness correlates strongly with phenotypic similarity.

Given the often complex associations between genotype and phenotype, recent work has often leveraged machine learning to produce intricate models trained on empirical data. Traditionally, the focus has been on feature-based approaches, using genetic annotations from which phenotypes are inferred \citep{Wood2014-tw,doi:10.1128/msystems.01045-20,wood2019improved}.
For example, Traitar \citep{weimann2016genomes} uses support vector machines with a sparsity penalty to predict phenotypes based on Pfam annotations \citep{mistry2021pfam}.\footnote{There is a wide array of resources and platforms for computational microbiome research, such as the MGnify platform for microbiome sequence data analysis \citep{richardson2023mgnify}, SPIRE for searchable database integrating diverse information derived from metagenomes including many modalities \citep{schmidt2024mode}, online analysis platforms \citep{alam2021kaust}, and more traditional protein family databases/mappers like NMPFamsDB \citep{baltoumas2024nmpfamsdb} or eggNOG \citep{cantalapiedra2021eggnog}.}
Those features can be aggregated over large collections of genes to use as input for machine learning methods \citep{doi:10.1128/msystems.00101-16,10.1093/bioinformatics/bty928,10.1371/journal.pgen.1007333,hernandez2022machine,d2023advancing}.
A different approach is to ignore gene-level information and directly work on taxonomic compositional count data \citep{li2015microbiome,calle2019statistical,knight2018best,zhou2019review,huang2023supervised}.
\citet{djemiel2022inferring} provides a high-level overview of existing work on functional inference from microbiota.

Despite the impressive progress achieved by these efforts, recent advances suggest that incorporating long stretches of genome sequences can enhance our understanding of genotype-phenotype relationships \citep{eraslan2019deep,alharbi2022review,deschenes2023gene,hammack2023machine}. Deep learning applied to raw DNA data, such as CNNs for taxonomy prediction \citep{rojascarulla2019genet} or unsupervised training of transformers on k-mers as tokens \citep{Ji2020.09.17.301879}, has indeed shown promise in this regard, offering a more nuanced view of the genetic underpinnings of complex phenotypes. Recently, several ML-based methods have also offered to prioritize non-coding variants; still, the recognition of disease-associated variants in complex traits, such as cancers, is challenging \citep{alharbi2022review}. On a methodological level, operating on (collections of) entire genomes at the sequence level remains difficult \citep{alharbi2022review}. Even recent approaches to scale transformers to longer sequences via linear attention models \citep{dai2019transformerxl,sukhbaatar2019adaptive,rae2019compressive,child2019generating,beltagy2020longformer,zaheer2021big} or reducing sequence lengths up front by stacked shifted window transformers \citep{liu2021swin} cannot directly be scaled to entire (collections of) genomes.

\begin{figure*}
    \centering
    \includegraphics[width=0.9\textwidth]{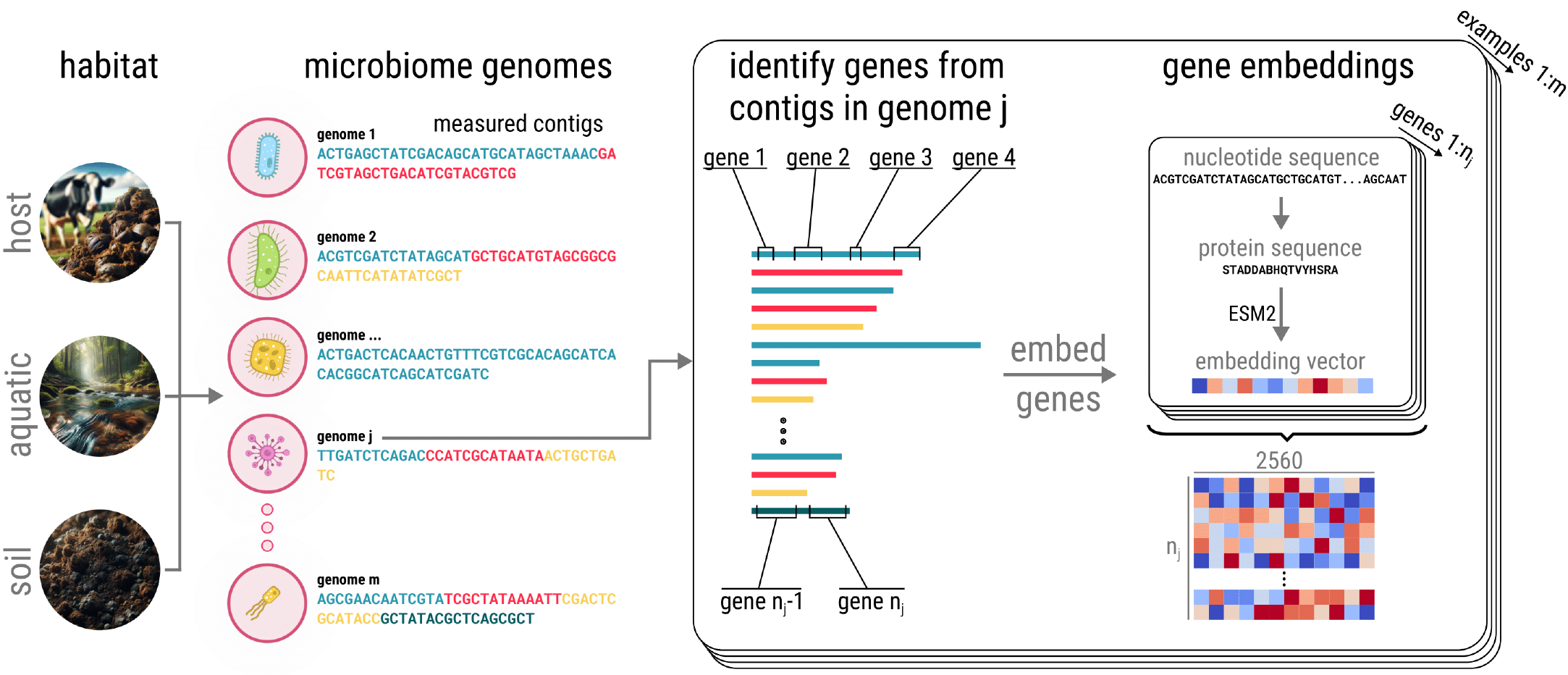}
    \caption{A conceptual overview of our data preprocessing pipeline. Each sample stands for an entire genome, reconstructed from shotgun sequencing in terms of contiguous consensus regions (contigs). We identify all genes within each contig (using Prodigal) and embed the corresponding protein sequences using an existing protein large language model (ESM-2) into a $\demb$-dimensional vector space.
    A single `input example', corresponding to an entire genome, is ultimately represented by a $(n_j \times \demb)$-dimensional tensor.}\label{fig:overview_data}
\end{figure*}

\xhdr{Contributions.} In this work, we focus on habitat specificity, i.e., predicting and understanding a complex phenotype---where a microbiome sample was collected---directly from the collection of genomes of the organisms in the sample.
First, we phrase this as a classification task via the following steps: (a) comprehensive gene identification from a sample of genomic sequences, (b) compute fixed-size vector representations for each gene using the multi-billion parameter ESM-2 model \citep{lin2023evolutionary}, (c) train our encoder-only transformer on sequences of gene embeddings with a classification head for the habitat.

Next, we develop attribution techniques to extract highly habitat-predictive pairs of genes by: (a) retrieving pairs of genes with high attention scores, (b) clustering pairs by similarity, (c) looking up the genes from pairs within a cluster in existing databases \citep{10.1093/molbev/msab293}, and (d) constructing gene interaction networks for individual genomes.
We train our model on a large subset of ProGenomes v3, a dataset of almost 1 million high-quality prokaryotic genomes \citep{mende2020progenomes2,fullam2023progenomes3}.

Our empirical evaluations provide multiple insights.
(a) Given the complexity of the phenotype, we obtain strong classification performance.
(b) Our attribution is among the first to assess the importance of gene co-occurrence across entire genomes for phenotype prediction. It recovers some known interactions, and we hypothesize that it proposes good candidates for experimental follow up.
(c) Our findings indicate that exploiting sequence level information is beneficial compared to functional or taxonomic annotations when predicting phenotype from genotype---in line with recently stated conjectures \citep{deschenes2023gene,hammack2023machine}.
In summary, studying how interactions among large collections of genes/proteins relate to complex phenotypes (such as habitat) directly from sequence level data holds great promise to advance our understanding of how the microbiome interacts with hosts and environments alike.
While we focus on habitat specificity, we highlight that our methodology is not limited to such broad classification tasks from microbial data, but extends to other tasks and domains.

 \begin{figure*}
    \centering
    \includegraphics[width=0.32\textwidth]{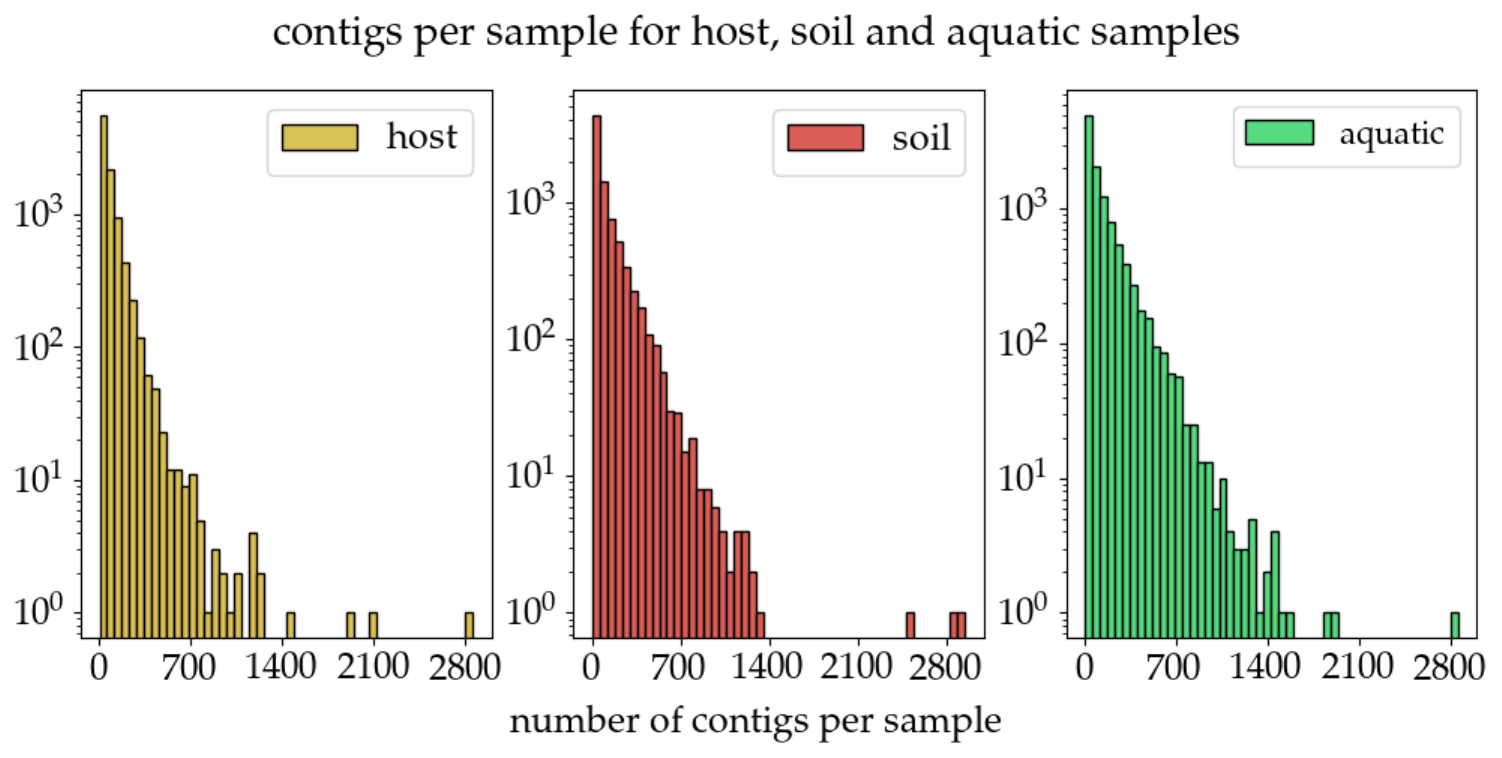}\hfill
    \includegraphics[width=0.32\textwidth]{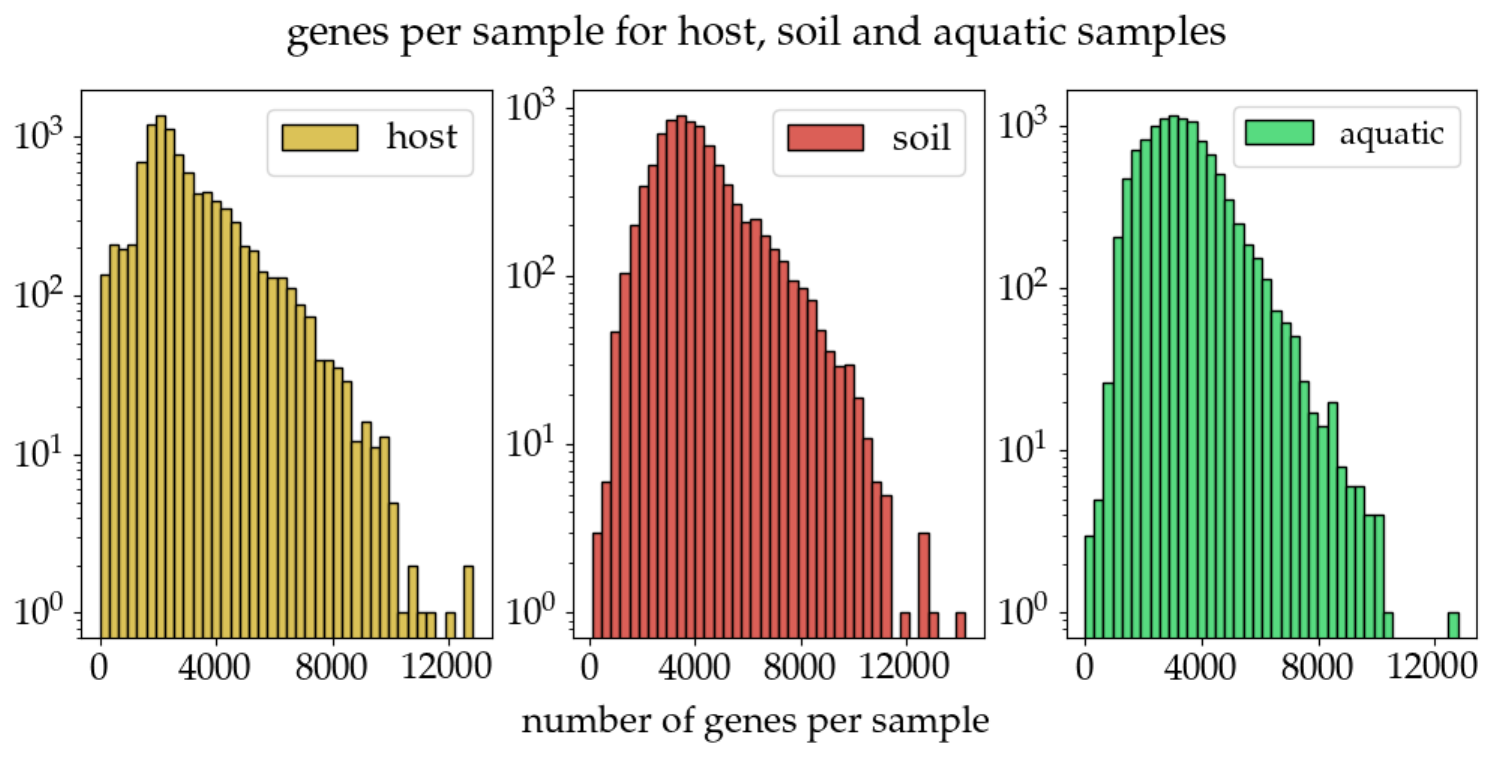}\hfill
    \includegraphics[width=0.32\textwidth]{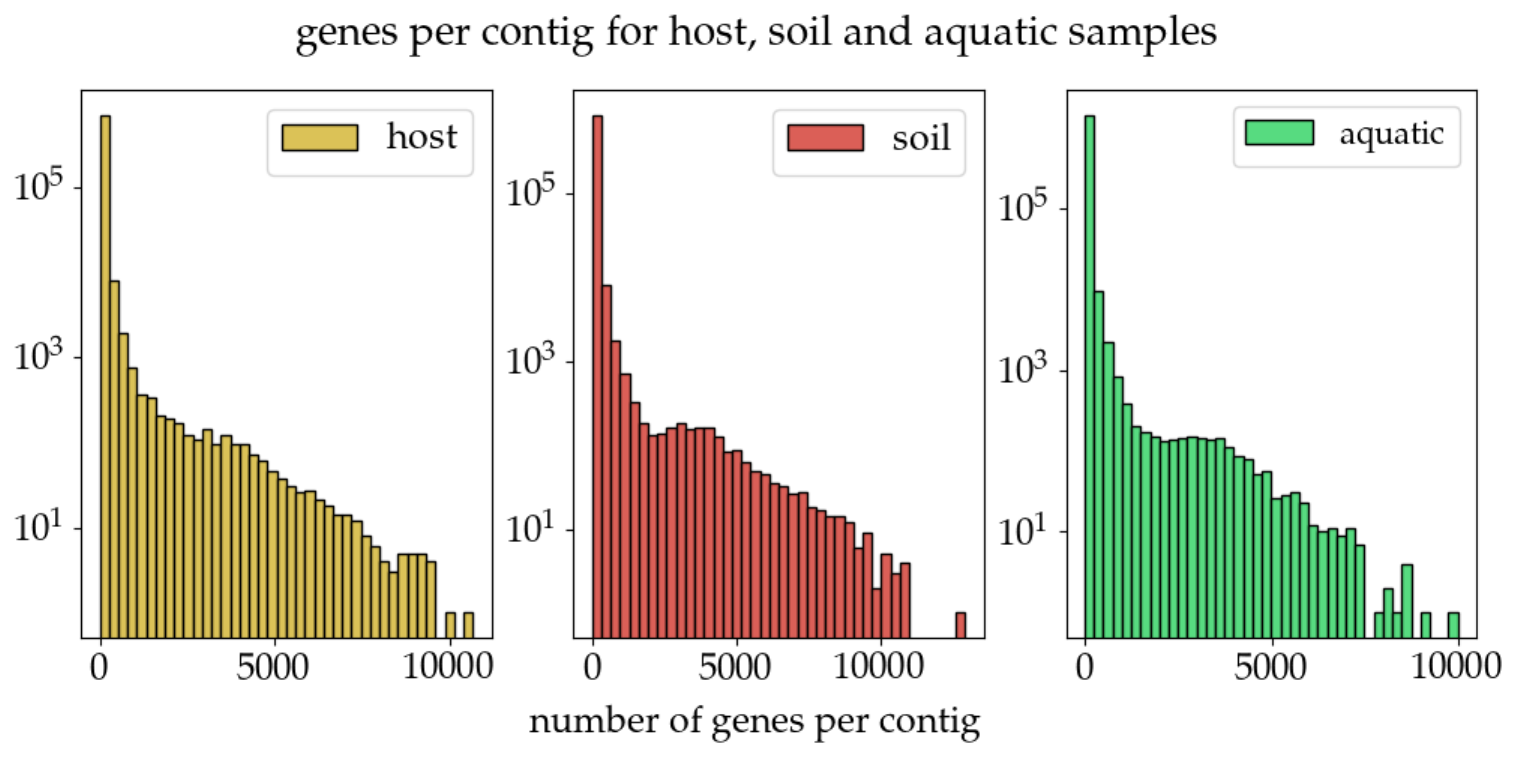}
    \caption{\textbf{Left:} Histogram of the number of contigs per sample (genome). \textbf{Center:} Histogram of the number of genes per sample (genome). \textbf{Right:} Histogram of the number of genes per contig.}\label{fig:data_stats}
\end{figure*}

\section{Methodology}\label{sec:method}

\xhdr{Microbiome data.}
Various peculiarities arise from the prevailing sequencing technology \cite{Ghurye2016-pk} used for large scale microbial DNA sequencing screens as collected by ProGenomes \citep{Mende2016-hl,10.1093/nar/gkz1002,fullam2023progenomes3}.
For example, instead of obtaining entire genomes, one typically only reconstructs so-called `contigs', i.e., contiguous consensus regions of DNA that have been recovered from the short sequenced snippets.
While different chromosomes are expected to produce different contigs, even circular, single-chromosome genomes may lead to multiple contigs.
While genes appear in the right order within a contig, we typically cannot determine the order in which contigs appear within the full genome.
We limit our attention to coding genes, requiring us to identify individual genes from within each contig.
Our tailored data-preprocessing aims at accounting for these task-specific aspects.
\Cref{fig:overview_data} provides an overview of the first stage of our framework.

\xhdr{Dataset.} We obtain all genomic data from ProGenomes v3, an open-source database comprising over 900,000 consistently annotated bacterial and archaeal genomes from over 40,000 species.
Collectively, the genomes contain 4 billion genes; for reference, the human genome contains about 20,000 coding genes.
Consistent phenotypic data across all genomes in the database is limited, so we focus on habitat classification in order to comprehensively utilize the available genomic data and assess prediction performance for a complex phenotype.
We select the three habitats with the most associated genomic data: \emph{host} (symbiotic or parasitic microbiome, which relies on a host organism, typically collected from animal feces), \emph{soil} (generally free-living microbiome collected from the soil), and \emph{aquatic} (free-living microbiome collected from natural water bodies). In total, our genome dataset comprised $m = 29,089$ genomes (soil: 8,248; host: 9,770; aquatic: 11,070) and 3,056,557 contigs with a mean length of $3445 \pm 1632$ genes.

\xhdr{Gene embeddings.} The high variability of contig lengths in our dataset challenges the direct application of existing deep learning approaches.
We therefore deploy a multi-gene approach that leverages an existing protein large language model to produce fixed-sized embeddings as input to our model.
Our workflow consists of identifying coding genes in each contig with Prodigal \citep{Hyatt2010}, which results in $33 \pm 179$ genes per contig.
\Cref{fig:data_stats}(left) shows the distributions of how many contigs are contained in a sample with a clear skew towards few contigs per sample (note the logarithmic y-axis), \cref{fig:data_stats}(middle) shows the distribution of the overall number of genes extracted per sample, and \cref{fig:data_stats}(right) shows the distributions of genes per contig, which is also heavily skewed towards small contigs.
The common peak at around 4,000 genes per sample aligns well with expectations of average gene counts in bacteria and archaea.
We then use ESM-2 (3B) \citep{lin2023evolutionary} to embed each amino acid sequence identified by Prodigal into a fixed-dimensional ($\demb = 2560$) vector space.
Ultimately, for each sample $j$ (i.e., each genome) we stack all $n_j$ gene embeddings belonging to that sample into a $(n_j \times \demb)$-dimensional tensor, where $n_j$ still varies across samples and which comprises one `input example' for our model.
For the roughly 8k, 10k, and 11k samples from soil, host, and aquatic habitats (a total of $m=29,089$ training examples), respectively, this yields a total of almost 1TB of pre-computed ESM-2 gene embeddings as the final dataset for our transformer model.
\Cref{fig:overview_data} provides a conceptual overview of our data preparation process.

\begin{figure*}
    \centering
    \includegraphics[width=0.9\textwidth]{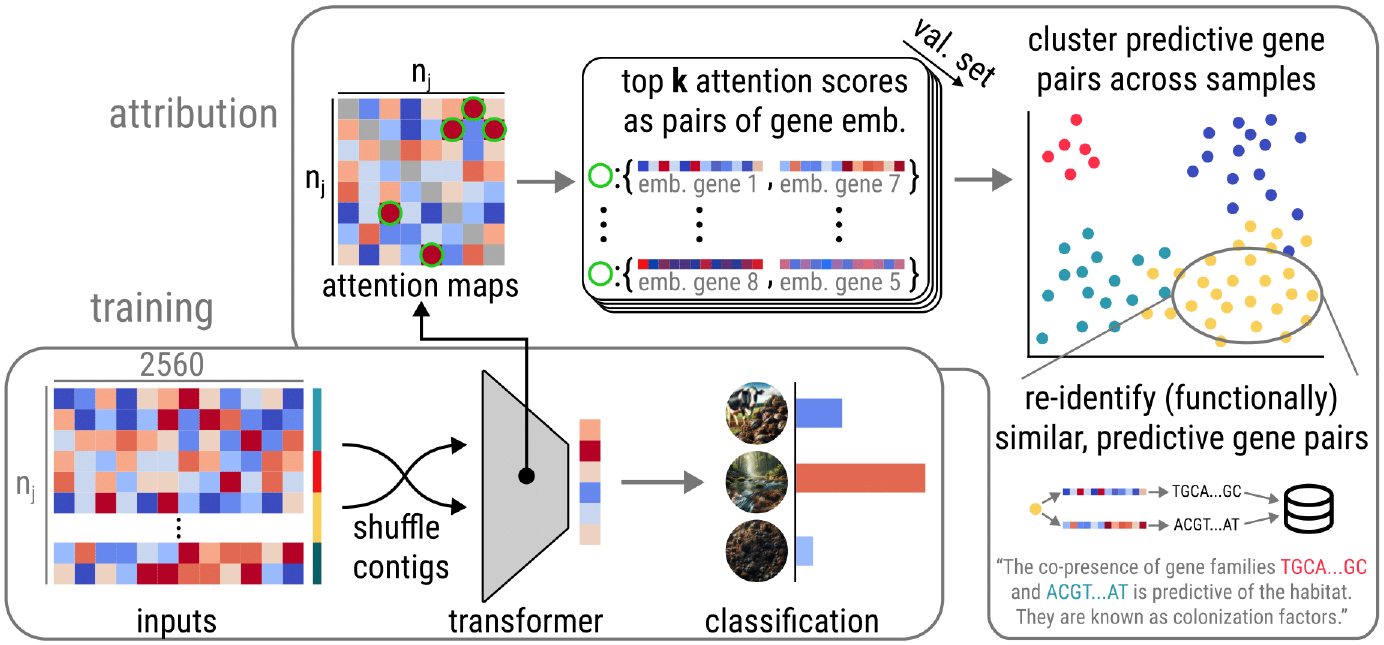}
    \caption{A conceptual overview of our training and attribution pipelines. \textbf{Training:} We feed the $(n_j \times \demb)$-dimensional inputs to our transformer, interpreted as a sequence of $n_j$ `tokens', each already represented by a fixed $\demb$-dimensional embedding. We randomly shuffle the contigs within each sample, since the `correct' order is unknown. Our model is then trained with the cross-entropy loss for classification.
    \textbf{Attribution:} After training, we extract the last-layer attention maps for all validation samples. We find the indices of the top-$k$ attention scores in each map, i.e., which gene embedding attends strongly to which other gene embedding. We cluster these pairs and visualize the clustering via non-linear dimensionality reduction. Within each cluster, we then re-identify the nucleotide sequences of all genes within all pairs and match them against gene annotation databases.}
    \label{fig:overview_training}
\end{figure*}

\xhdr{Model architecture and training.}
Since individual genes are typically shared by many organisms within and across habitats, we hypothesize that habitat specificity heavily depends on the co-presence and interaction effects of multiple genes.
For these interactions, the local context is relevant because functionally related genes tend to be clustered in local neighborhoods on the genome \citep{doi:10.1126/sciadv.aax6525}.
The attention mechanism in transformer architectures \citep{vaswani2017attention} is not only well suited to capture such associations in making predictions but also allows for attribution techniques to extract relevant pair-wise interaction effects (c.f., attribution paragraph in \cref{sec:method}).
Hence, we propose an encoder-only BERT-like architecture \citep{devlin2019bert} for classification (using the standard cross-entropy loss) with $15$ layers, a single attention head, and a hidden dimension of $640$.
To reduce the memory footprint during training, we feed the original embeddings of dimension $\demb = 2560$ obtained from ESM-2 into a single linear layer to obtain a reduced hidden dimension of $640$.

We set the maximum input sequence length to 4096, reaching beyond the average number of genes within a genome.
Because some samples in our dataset contain more genes than that (c.f., \cref{fig:data_stats}), we truncate them.\footnote{Note that each of these 4096 'tokens' represents an entire gene, each of which can consist of thousands of base-pairs. Therefore, the `effective' context window approaches $10^7$ base-pairs.}
Here, we make use of the fact that the order of genes is preserved within contigs, but not across contigs. Specifically, in each epoch, we randomly permute the contigs within every input example before potentially truncating (c.f., \cref{fig:overview_training}).
Over multiple epochs, this procedure allows the model to learn dependencies between all possible pairs of genes even for the longest examples despite the limited maximum sequence length. 
Moreover, the permutation may encode our prior knowledge that there is no intrinsic (known) order among the contigs within an example as an invariance in the model.
While various techniques for sparse and/or linear attention \citep{tay2021long} may allow us to extend the maximum input sequence, it would impede attention-based attribution, as we would not obtain comparable attention scores for all pairs of genes.
Similarly, recent techniques scaling transformers to millions of base pairs such as Hyena \citep{nguyen2023hyenadna} rely on dilated convolutions on the input sequence, rendering attribution to interactions difficult. 
Therefore, we opted for full attention using FlashAttention \citep{dao2022flashattention} during training, which still allows us to extract complete attention scores during attribution/validation.

Overall, our model consists of over 68 million trainable parameters.
We used AdamW \citep{loshchilov2019decoupled} with linear learning rate decay for 16 epochs on 4 NVIDIA A100 (40GB) GPUs until convergence of the out-of-sample performance on the validation set.

\xhdr{Attribution techniques.}
During training, we hold out $\nval = 1453$ samples for validation and our attribution analysis.
The goal of our attribution technique is to extract gene-pairs or even larger collections of genes whose co-presence in a given sample is predictive of the habitat.
While genes within a pair need not necessarily physically interact as in protein complexes, we posit that they `interact' in being jointly specific to the habitat.
We propose the following procedure for attribution, which we depict in \cref{fig:overview_training}.
\begin{enumerate}
    \item For each sample in the validation set (each consisting of a collection of fixed-size gene embeddings grouped into contigs; c.f., \cref{fig:overview_training}) that was classified correctly with certain confidence (top softmax value above $0.85$), compute all last-layer attention maps and extract the positions (indices) of the top-k scores for a fixed $k \in \mathbb{N}$. Following common practice in the literature (starting with \citet{vaswani2017attention}), we interpret high attention scores as relevant to the prediction task. Each of the extracted $\nval \cdot k $ indices corresponds to a pair of input gene embeddings $\{p_i := (\emb^1_i, \emb^2_i)\}_{i=1}^{\nval \cdot k}$ for $\emb^j_i \in \bR^{\demb}$. 
    \item
    In this step, we use DBSCAN \citep{10.5555/3001460.3001507} as a clustering algorithm, which has the advantage of inferring the number of clusters by itself, and cluster via the following custom distance function
    \begin{align*}
        \mathrm{dist}(p_i, p_j) = \min\{ &2 - S_c(x^1_i, x^1_j) - S_c(x^2_i, x^2_j),\\
                                         &2 - S_c(x^1_i, x^2_j) - S_c(x^2_i, x^1_j)\}\:,
    \end{align*}
    where $S_c$ is the cosine similarity and we are agnostic about the order of the genes within the pair.
    \item For each point $p_i$ in each cluster, we recover the two gene sequences that produced the gene embeddings $\emb^1_i, \emb^2_i$.
    We then perform sequence similarity search on all these genes in the databases EggNOG \citep{cantalapiedra2021eggnog}, KEGG orthologs \citep{10.1093/nar/gkv1070}, and NCBI Blast \citep{Altschul1990-mm, Boratyn2019-ds,Camacho2023-lw} to extract functional and taxonomic annotations.
    \item We propose gene interaction networks loosely inspired by gene pathways. If a certain gene appears in more than one of the pairs \emph{within a sample}, we use these overlaps in the extracted $k$ pairs of genes to construct a gene network. Genes that are hubs in these networks have many highly predictive interactions with other genes and may thus be of particular functional importance.
\end{enumerate}

\section{Results}\label{sec:results}

\xhdr{Why habitat classification?}
The reason we focus on the seemingly `simple' three-way classification of habitats (host, soil, and aquatic) is three-fold. First, habitat is a broad and highly complex phenotype, which is difficult to predict directly from genotype. Hence, strong performance on this task indicates that our general framework may apply equally to other phenotypes. Second, it is straightforward to compare feature attributions among all three classes in order to help validate our approach. Third, habitat annotations are typically reliable and widely available for microbiome samples.
In the remainder of this section, we particularly focus on extensive internal and external validation results demonstrating that our modeling approach manages to pick up on the importance of the co-presence of genes.

We conjecture that gene pairs (or collections/networks) found by our attribution technique are of biological interest in various ways.
For example, when predicting host-related habitats, such gene clusters may shed light not only on specific genes but also on gene interaction networks that may be involved in colonization \citep{stephens2015identification,powell2016genome,kemis2019genetic}.
When the identified gene pairs are found in gene annotation databases and have known functional annotations, we can directly point to interactions of functional aspects associated with the predicted phenotype and potential colonization properties.
On the contrary, when the found genes are part of the ``microbial functional dark matter'', we hypothesize they are good candidates to follow up on experimentally.
For example, one could knock out the predicted genes and measure the abundance of the mutant versus wild type in a model habitat \citep{doi:10.1073/pnas.1610856113, doi:10.1073/pnas.1014971108, Brouwer2020-xq}. 

\begin{table}
\centering
\caption{One-vs-rest performance of our model.}\label{tab:combined}
{\small
\begin{tabularx}{0.8\columnwidth}{*{5}{C}}
\toprule
\textbf{class} & \textbf{samples} & \textbf{precision} & \textbf{recall} & \textbf{F1} \\ 
\midrule
\rowcolor{gray!20}\multicolumn{5}{c}{\textbf{test set}} \\
\midrule
host & 488 & 0.84 & 0.80 & 0.82 \\ 
soil & 412 & 0.63 & 0.43 & 0.51 \\ 
aquatic & 553 & 0.66 & 0.84 & 0.74 \\
\midrule
\rowcolor{gray!20}\multicolumn{5}{c}{\textbf{pseudo-samples}} \\
\midrule
host & 488 & 0.58 & 0.82 & 0.68 \\ 
soil & 412 & 0.58 & 0.16 & 0.24 \\ 
aquatic & 553 & 0.58 & 0.69 & 0.63 \\
\bottomrule
\end{tabularx}
}
\end{table}

\xhdr{Classification performance.}
We evaluate our model on $\nval = 1453$ held out samples from the ProGenomes v3 dataset.
It achieves an overall accuracy of $71\%$ (\cref{tab:baseline}).
Given the complexity of the task (see \cref{sec:intro}), this is a strong performance for our 3-way classification task. For more detailed comparisons with the baselines, we refer reader to the \cref{app:baselines}.
\Cref{tab:combined}(top) shows how performance varies across habitats: while host samples are identified well, samples from the soil are often misclassified as aquatic.
Biologically, host microbiomes are mostly symbiotic or parasitic, where they tend to lose unneeded portions of their genome due to deletional bias in bacterial genomes \citep{McCutcheon2012-hz, Boscaro2017-xb}.
This arguably leads to substantial genomic differences from free-living microbiomes in soil or aquatic environments, which conversely can have strong adaptability due to their versatile metabolic pathway and, therefore, can survive in a variety of environments \citep{Shu2022-wi,doi:10.1126/science.adf4444}.
There is likely also a more direct mixing of microbiomes inhabiting soil and aquatic environments, rendering distinguishing soil from aquatic examples incredibly difficult.
Finally, the sample imbalance in our training set is slightly skewed towards aquatic examples.
\Cref{app:ablation} provides an ablation of how the number of layers, size of feedforward layers, and embedding dimension affect model performance.

\xhdr{Internal validation.}
To provide some internal validation of the effectiveness of our attribution technique, we construct `pseudo-examples', inputs to our model that consist only of genes that were identified by the attribution to be part of highly-predictive pairs for $k=100$.
We randomly concatenate the respective gene embeddings from each validation example (without repetitions) to form `pseudo-examples' which consist on average of only about $100$ genes.
These pseudo-examples (a) present only about $3\%$ of the original genomes, and (b) only serve as a bag of genes in that the true order of genes on the genome (or within contigs) is lost---typically crucial information \citep{Salaverria2011-hs}.
Given those limitations, we would expect classification performance to drop to essentially random guessing unless the genes contained in the `pseudo-examples' are highly predictive for the habitat.
Our model still achieves an overall accuracy of $58\%$, substantially better than random guessing.
\Cref{tab:combined}(bottom) shows that the model can still extract useful information from host and aquatic `pseudo-examples'.
This provides strong evidence that gene pairs identified by our attribution, indeed contain a significant number (and important combinations) of habitat-specific genes. 

\xhdr{Clustering.}
The purpose of running gene pair clustering is two-fold: it serves as additional validation that allows us to judge the consistency of gene pair prediction across all genomes from the same environment. At the same time, it can provide us with new perspectives on understanding the function of genes and the relationship between genotype and phenotype. 
We expect gene pairs within the same cluster to have similar functions, and that a cluster reflects common gene families shared by microbes from a given habitat.

\begin{figure*}
    \centering
    \includegraphics[width=0.32\textwidth]{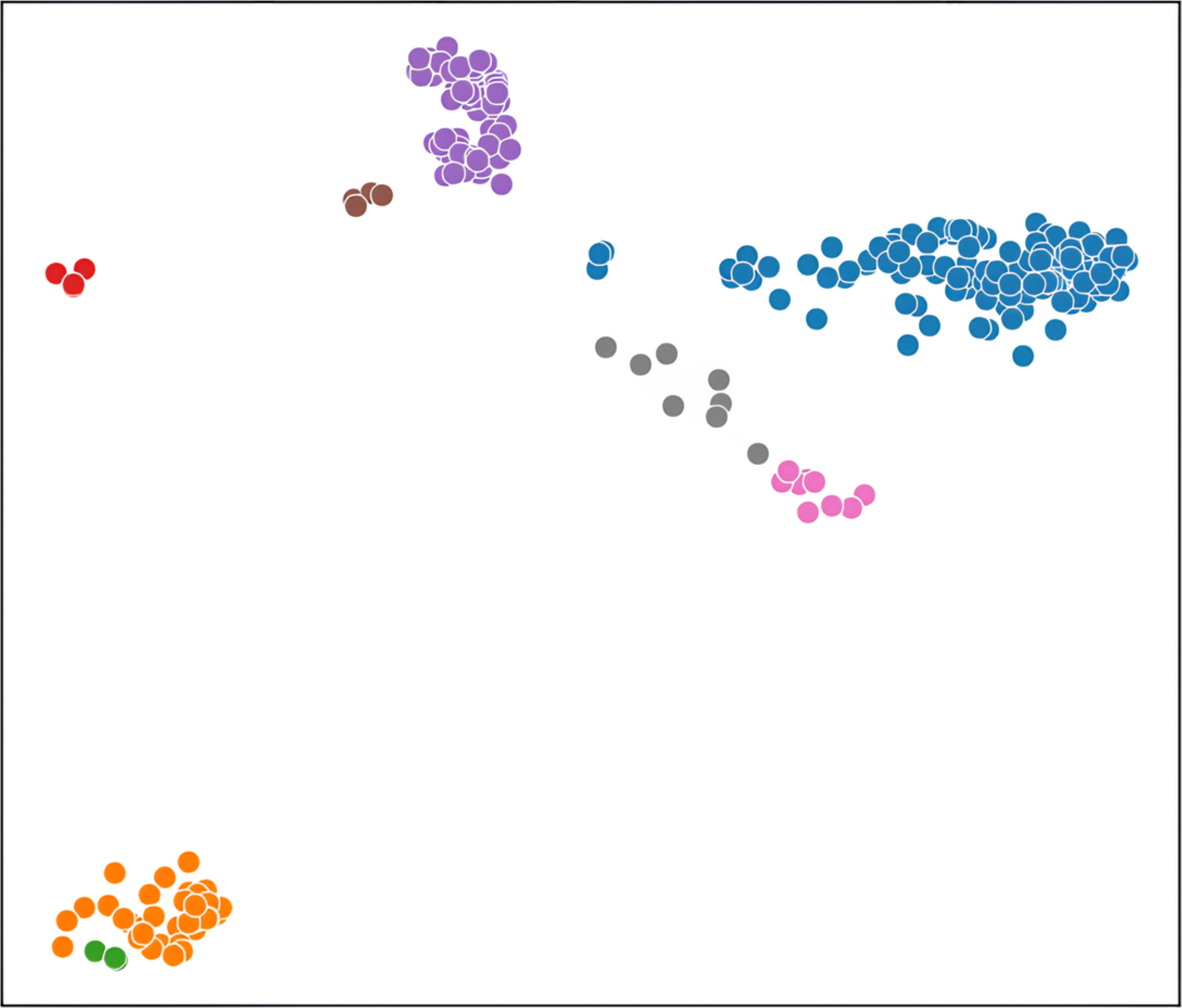}\hfill
    \includegraphics[width=0.32\textwidth]{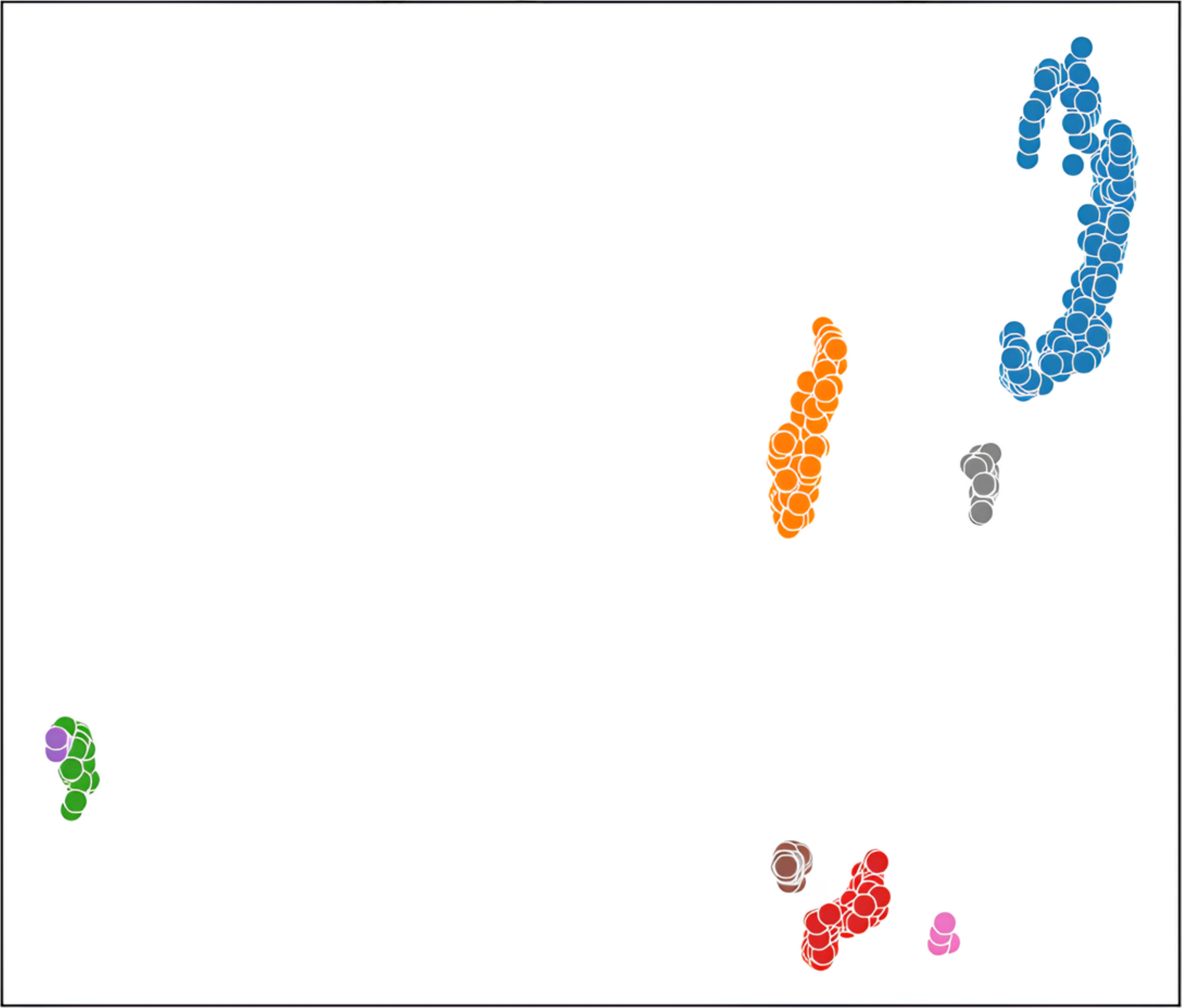}\hfill
    \includegraphics[width=0.32\textwidth]{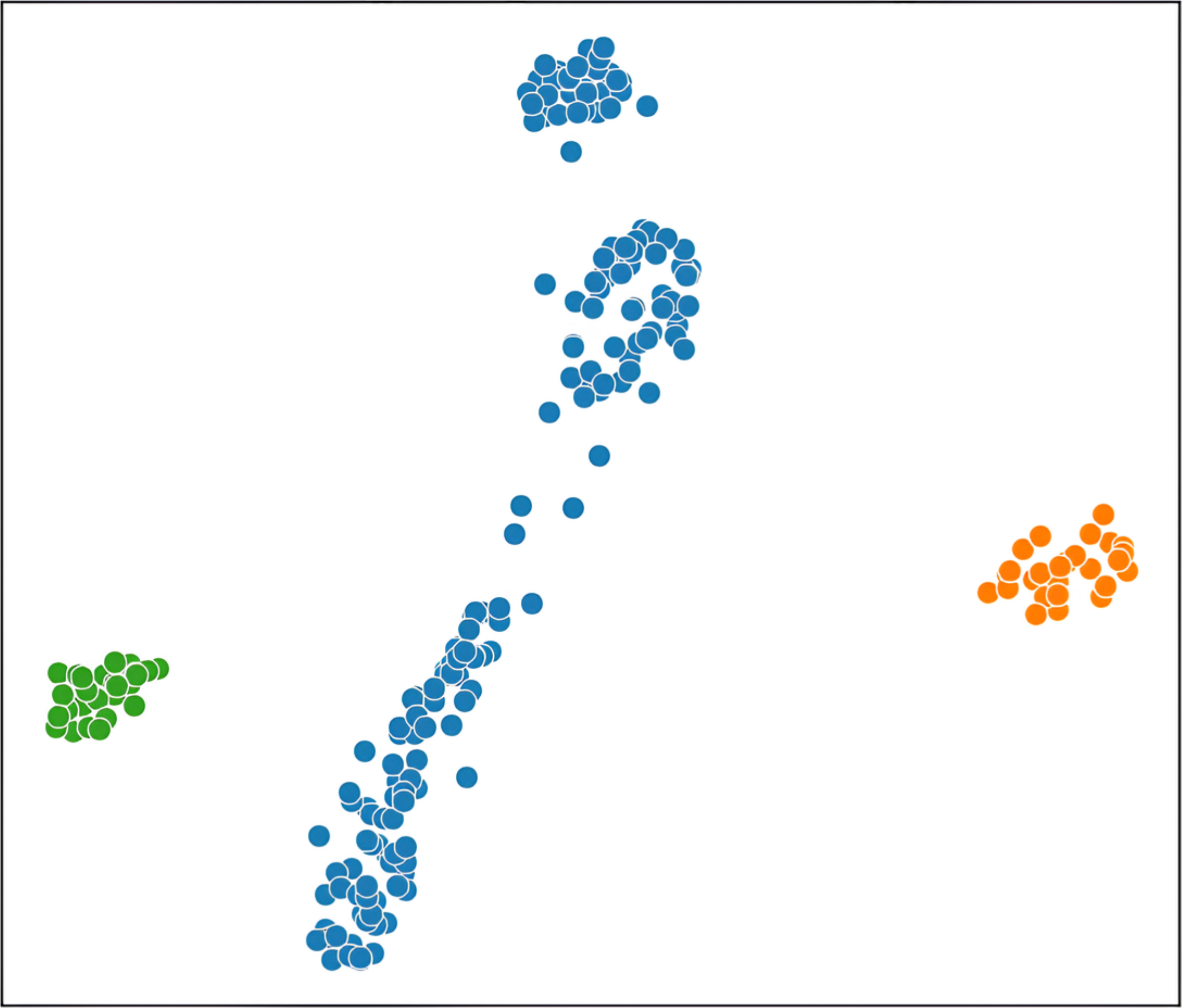}
    \caption{Two-dimensional visualization of the clusters for aquatic (left), host (middle), and soil (right) samples via UMAP \citep{mcinnes2018umap}, omitting points not belonging to any cluster.} \label{fig:umaps}
\end{figure*}

In \cref{fig:umaps} we illustrate the gene pair clusters using UMAP \citep{mcinnes2018umap}.\footnote{We omit `outliers', i.e., points that did not belong to any cluster after DBSCAN finished for a clearer illustration. These outliers are bound to exist due to the breadth of habitat as a phenotype.}
The pairs of genes cluster well, indicating that gene pairs within a cluster are functionally similar as measured by the distance of their ESM-2 embeddings.
Further, different clusters are well-separated, indicating that we have identified different `hubs' of gene interactions that are individually predictive of the habitat.
For completeness, we provide similar plots using t-SNE \citep{JMLR:v9:vandermaaten08a} instead of UMAP in \cref{fig:tsnes} in \cref{app:visuals}, showing that the clear separation of clusters is not specific to the choice of dimensionality reduction technique.

We further verified that within most found clusters, gene families are quite uniform.
From the extracted functional and taxonomic annotations, we found that the clusters recover biologically plausible `functional factors'. 
For example, in the largest (blue) cluster from host samples, most of the pairs share the KEGG orthologs \citep{10.1093/nar/gkv1070} K01992 and K11051.
The latter is known as multidrug/hemolysin transport system permease, a protein that plays an important role in bacterial infection of animal hosts.
In the largest (blue) cluster from aquatic samples, most gene pairs share the K08226 functional ortholog.
Genes from this ortholog code chlorophyll transporter.
This matches our knowledge that most photosynthetic bacteria, such as Cyanobacteria and Chlorobi, live in water.
In the largest (blue) cluster from soil samples, we found the following frequent orthologs: K01535, K01531, K17686, K01533, and K17686. These gene families are all involved in ion transport.
For completeness, we provide all found orthologs in all of the clusters for the three classes in \cref{app:orthologs}.
Great care must be taken when associating biochemical functions of single-gene coded proteins with complex phenotypes.
However, we believe that surfacing interpretable pointers toward potentially relevant interactions from full genome data is a promising tool to guide hypothesis formation for experimental colonization studies.

\begin{figure}
    \centering
    \includegraphics[width=0.9\columnwidth]{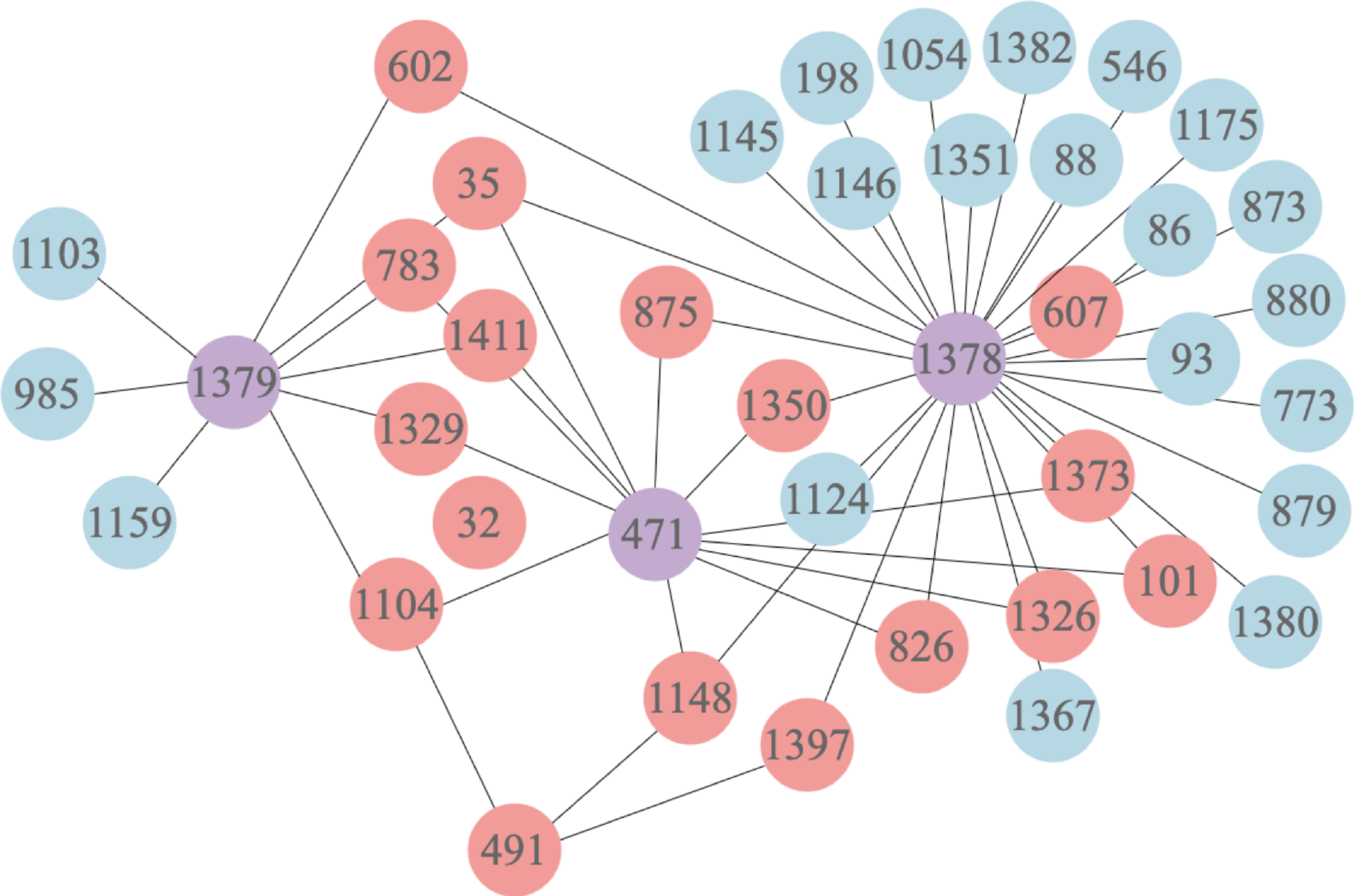}
    \caption{Gene interaction network for the sample 1311.SAMN14644158; coral/violet: genes with more than one neighbor (hubs); blue: genes with one neighbor (peripheral); Purple hubs are described in the text. Numbers are in order of appearance on the genome.}\label{fig:pathway}
\end{figure}

\xhdr{Gene interaction networks.}
We present an example of one of the gene interaction networks constructed by our attribution technique in \cref{fig:pathway}.
The genome from which this network was constructed belongs to Streptococcus agalactiae, a commensal bacterium.
Although it colonizes the gastrointestinal and genitourinary tract of up to $30\%$ of healthy human adults, it is still poorly understood.
We could only find functional annotations for 14 of the 41 genes in the network. The rest of the genes have no annotation via our methodology.
In particular, the gene with the most connections, gene 1378, is identified as a peptidoglycan bound protein that can have various functions, including roles in cell wall synthesis, cell division, and interaction with the environment.
In the context of bacterial colonization, peptidoglycan-bound proteins can contribute to the adherence of bacteria to host tissues, evasion of the host immune response, and establishment of infection \citep{10.1371/journal.pgen.1004433}.
Further, gene 1379, another highly connected hub in our network, is involved in dextransucrase activity. Dextransucrase is an enzyme that catalyzes the formation of dextran, which can contribute to the formation of biofilms, which are communities of bacteria that adhere to surfaces.
Biofilms play a crucial role in bacterial colonization, as they can protect bacteria from environmental stresses and enhance their survival and growth \citep{10.3389/fmicb.2019.00959,Lee2015-yg}.
Finally, gene 471, yet another highly connected hub, belongs to peptidase S8 family 5, also known as subtilases.
This enzyme plays important roles in colonization, including the degradation of host tissues and evasion of the host immune system \citep{plants12020369}.

These examples of gene annotations demonstrate our model's capability to predict not only habitat-specific genes but also how and with which other genes interact to become highly predictive of the habitat.
We hope to demonstrate with this example how our framework could be used by biologists to investigate concrete scientific questions around the relevance of gene interactions in complex phenotypes.
Further, we highlight that besides confirmatory evidence, our model can also be used to extract highly connected hubs across a large number of samples that are not found in existing databases, i.e., that are part of the `functional microbial dark matter'.
Such genes may be particularly well suited for experimental study in the quest of uncovering microbial dark matter.
In \cref{app:gin}, we also present examples of gene interaction networks for the other two habitats.

\section{Discussion and outlook}\label{sec:conclusion}

\xhdr{Summary.} We introduced a model predicting complex phenotypes, such as habitat, from entire genomes on the sequence level of microbial sequencing data.
Our attribution technique extracts pairs (and collections) of genes whose co-presence is highly predictive of the phenotype.
We train our model on high-quality prokaryotic genomes from ProGenomes v3 and demonstrate state of the art classification performance.
Internal and external validations evidence the usefulness of our method in uncovering habitat-specific gene pairs and generating interpretable gene interaction networks that can serve as powerful hypothesis generators.

\xhdr{Limitations and future work.} Our method handles large effective context lengths while preserving genes as meaningful input units.
However, the context length is ultimately still memory limited due to full attention computations.
Moreover, in line with our attribution goals, our analysis is limited to the coding regions of genomes.
Assessing how non-coding regions affect classification and attribution is an interesting direction for future work.
ESM-2 has primarily been trained on eukaryote proteins.
While our results indicate that ESM-2 still provides informative embeddings for prokaryotes, further analysis is required to assess whether existing models are ``general purpose'' enough to capture microbial diversity.
We believe that training similar foundation models specifically for microbiome research is a worthwhile endeavor.
Finally, we only used habitat as a phenotype.

Other directions for future work include applying our general framework to more fine-grained classification tasks such as predicting host range \citep{ji2023hotspot}, geographic distributions, virulence, or industrially important metabolic products.
For example, when predicting antimicrobial resistance, our attribution may uncover gene networks involved in developing certain types of antimicrobial resistance.
Ultimately, experimental follow ups are required to confirm the potential impact of our hypothesis generator on biological practice.
Finally, while not necessarily novel, we believe the broader framework of representing variable length collections of variable length sequences by replacing inner sequences via fixed-size embeddings from large sequence models holds great promise for future multi-omics data analysis.
Due to the potential ramifications of computational health research, we describe the broader potential impact of this work in \cref{app:impact}.

\appendix

\section{Baselines}\label{app:baselines}
Since our modeling approach is the only of its type, there exist large gaps in terms of data types, capability, and interoperability between existing works and ours. \Cref{tab:comparison} in \cref{app:modelcomparison} provides a detailed comparison to existing methods, highlighting in which way most of them fall short in our problem setting.
To compare raw predictive performance, we put together a strong baseline using k-mer counts as features with traditional machine learning classifiers.
This is a widely popular and typically highly effective approach to supervised ML on sequence data \citep{Dubinkina2016-vz, benoit2016multiple, Wood2014-tw}.
It avoids the necessity of annotations (highly incomplete for prokaryotes) and scales well to entire genomes.
The best performing traditional ML models in the literature on k-mer counts and in bioinformatics more broadly are often random forests \citep{10.1093/bioinformatics/btad662, 10.1371/journal.pgen.1007333} and SVMs \citep{doi:10.1128/msystems.00101-16, 10.1093/bioinformatics/btad662, 10.1093/bioinformatics/bty928}.
\Cref{tab:baseline} shows that our method achieves higher accuracy than these algorithms trained on k-mer counts for different typical values of k.
Finally, we highlight that by design (using k-mer counts as features), these methods cannot be interpreted in terms of a single gene or gene interaction importance.

\begin{table}
\centering 
\caption{Accuracy for random forests (RF) and SVMs using linear and RBF kernels.} 
\label{tab:baseline} 
{\small
\begin{tabularx}{\columnwidth}{@{}l*{3}{c}*{3}{c}*{3}{c}c@{}}
\toprule
& \multicolumn{3}{c}{RF} & \multicolumn{3}{c}{SVM linear} & \multicolumn{3}{c}{SVM rbf} & \textbf{ours} \\
\cmidrule(lr){2-4}\cmidrule(lr){5-7}\cmidrule(lr){8-10}
\textbf{k-mer} & 3 & 5 & 8 & 3 & 5 & 8 & 3 & 5 & 8 & -- \\
\midrule
\textbf{acc} & 57 & 58 & 59 & 57 & 62 & 56 & 63 & 67 & 68 & \textbf{71} \\
\bottomrule
\end{tabularx}
}
\end{table}

\section{Validation on the STRING database}
\begin{figure}
    \centering
    \includegraphics[width=0.54\columnwidth]{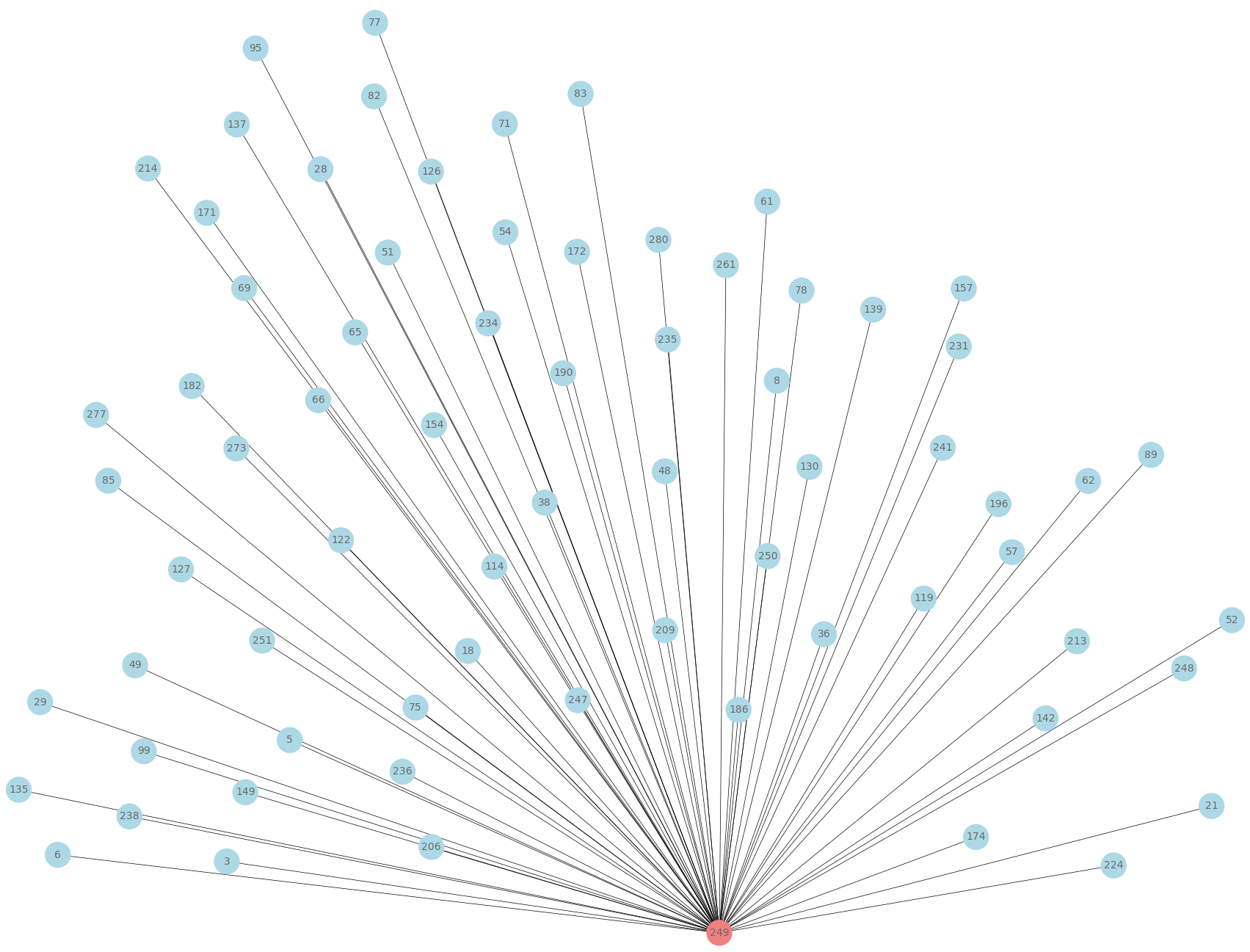}\hfill
    \includegraphics[width=0.4\columnwidth]{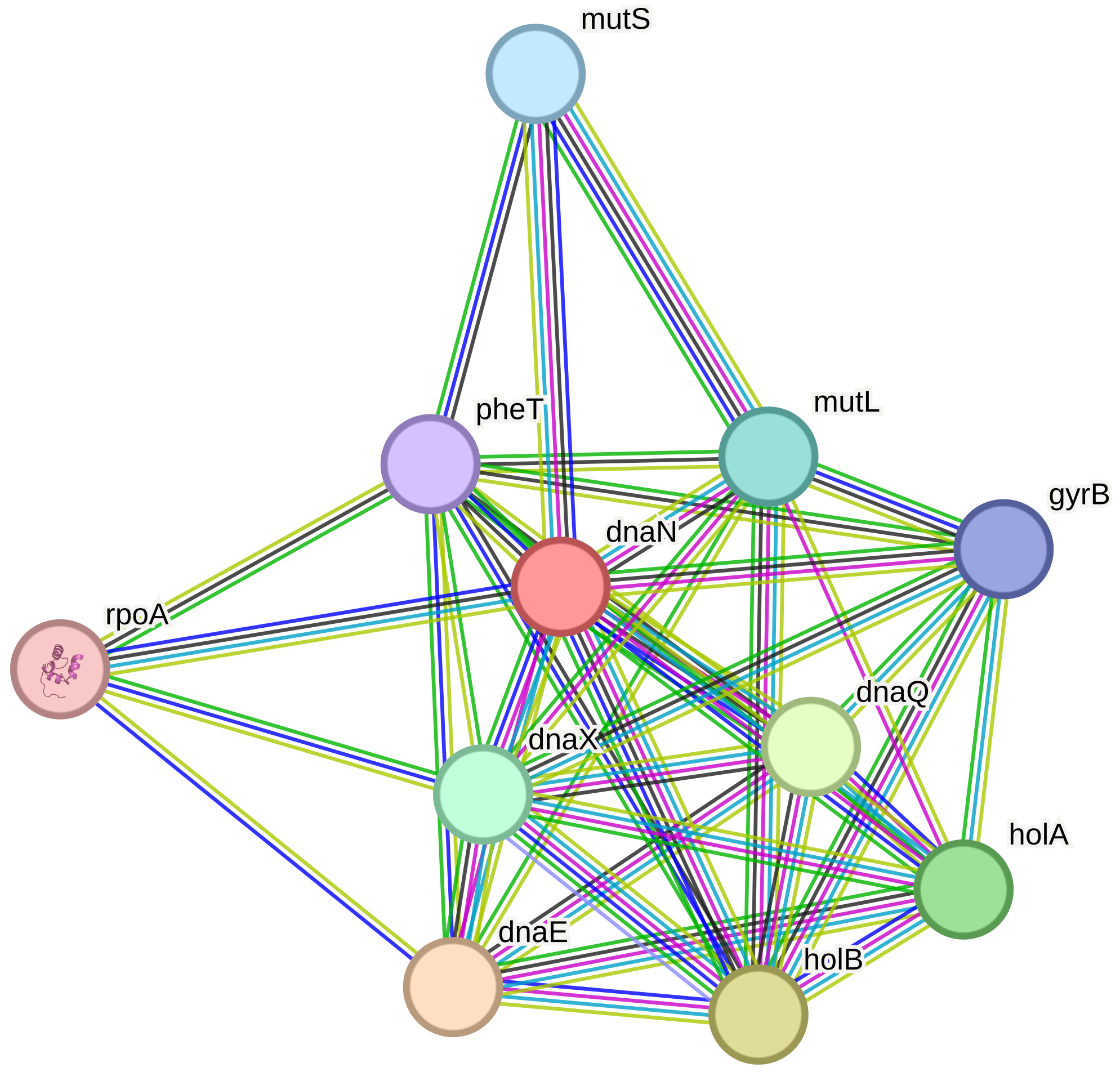}
    \caption{\textbf{Left:} Gene interaction network constructed for the sample 91844.SAMEA2820670. Coral color indicates genes with more than one neighbor (hub), while blue indicates genes with only one connection (peripheral). Genes are numbered by the order of their appearance on the genome. \textbf{Right:} Protein-protein interactions extracted from the STRING database \citep{Szklarczyk2023-qw} around dnaN. Only edges in magenta color are experimentally verified. Edges in other colors are predictions from the database.}
    \label{fig:pathway4}
\end{figure}

To further validate the biological relevance of our attribution technique, we turn to the STRING database \citep{Szklarczyk2023-qw}.
It systematically collects and integrates protein-protein interactions that contain both physical and functional associations.
Unfortunately, a majority of prokaryotic genes in our dataset are not found in the STRING database.
Hence, we validate our attribution on overlapping genes in our validation set and the STRING database.
We found an overlap for genes related to the survival of prokaryotes, such as DNA replication.
In sample 91844.SAMEA2820670, which is identified as Candidatus Portiera, our model identifies gene 249 as a hub.
This gene is annotated as DNA polymerase III beta subunit (dnaN), which is correctly found to interact with genes annotated as DNA polymerase III delta' subunit (gene 36, holB), DNA polymerase III epsilon subunit (gene 54, dnaQ) and type IIA topoisomerase (DNA gyrase/topo II, topoisomerase IV) B subunit (gene 250, gyrB), see \cref{fig:pathway4}(left).
The final DNA polymerase III is a result of pairwise interactions of the subunits.
In comparison, the similar (albeit more difficult to interpret) complex shown in \cref{fig:pathway4}(right) is obtained from the STRING database.

\section{Additional visualizations}\label{app:visuals}

\begin{figure*}
    \centering
    \includegraphics[width=0.32\textwidth]{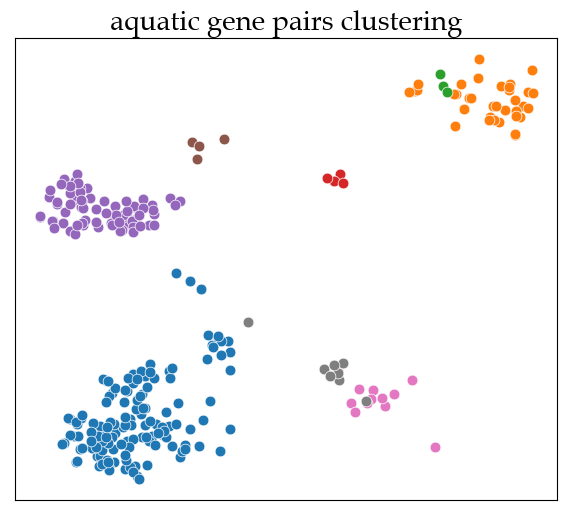}\hfill
    \includegraphics[width=0.32\textwidth]{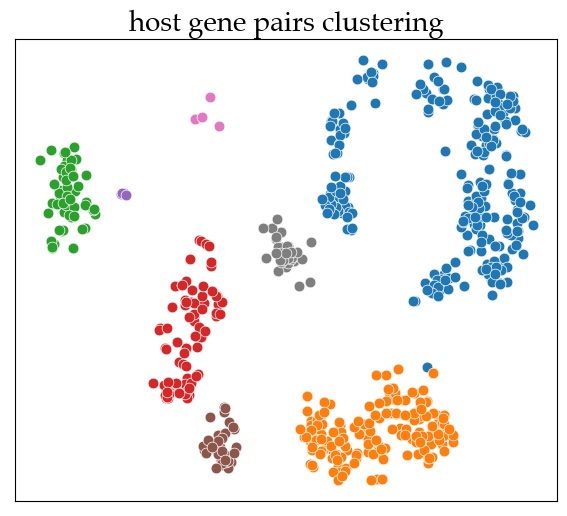}\hfill
    \includegraphics[width=0.32\textwidth]{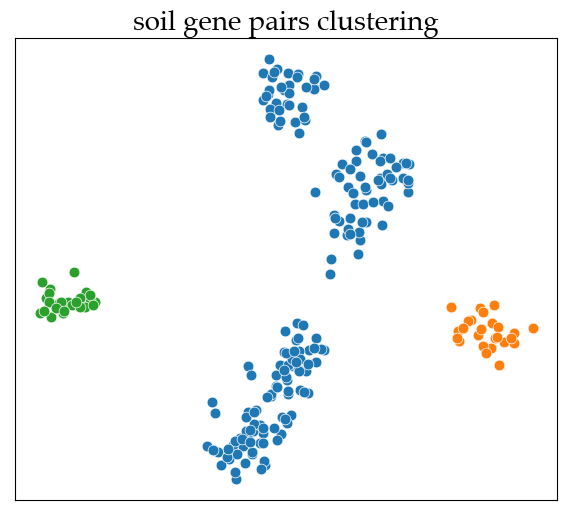}
    \caption{Two-dimensional visualization of the clusters for each of the three habitats aquatic (left), host (middle), and soil (right) separately via t-SNE \citep{JMLR:v9:vandermaaten08a}, where do not show points not belonging to any cluster.}\label{fig:tsnes}
\end{figure*}

Moreover, \cref{fig:tsnes} provides a visualization akin to the one in \cref{fig:umaps}, using t-SNE \citep{JMLR:v9:vandermaaten08a} for non-linear dimensionality reduction instead of UMAP \citep{mcinnes2018umap}. 
In both visualizations, the same clusters are clearly visible and separated, indicating the robustness of the found clusters to the specific dimensionality reduction technique.

\section{Ablation study}\label{app:ablation}

We now show ablations of varying the number of attention layers, feedforward layer size, and embedding dimension separately to understand their individual contribution to our model's performance.
\Cref{tab:ablation} compares different models in terms of overall accuracy as well as one-vs-rest classification metrics for each class.
These results highlight that downsizing the model in any way (10 instead of 15 layers, halving the feedforward layer size, or halving the embedding dimension using ESM2-650M instead of ESM-3B) drastically reduces performance roughly to the level of the baselines shown in \cref{tab:baseline}.
We highlight that the third and most impactful ablation of reducing the embedding dimension implies using a weaker embedding model, i.e., one may still achieve better performance by optimally compressing ESM-3B embeddings.
Overall, these results may indicate that our model is not yet ``larger than necessary'', i.e., one may still obtain moderate performance improvements by using an even larger model.

\begin{table*}
\centering 
\caption{Comparison of model configurations with different layers, layer sizes, and embedding dimension ($\demb$). The first model configuration has been used in this paper.} 
\label{tab:ablation} 
\begin{tabularx}{0.9\textwidth}{@{}C@{}C@{}C|@{}C@{}C@{}C@{}C@{}C@{}}  
\toprule
\textbf{layers} & \textbf{layer size} & \textbf{$\demb$} & \textbf{acc (\%)} & \textbf{class} & \textbf{precision} & \textbf{recall} & \textbf{F1} \\ 
\midrule
\multirow{3}{*}{15} & \multirow{3}{*}{2048} & \multirow{3}{*}{2560} & \multirow{3}{*}{71.2} & host & 0.84 & 0.80 & 0.82 \\
                    &                       &                       &                      & soil & 0.63 & 0.43 & 0.51 \\
                    &                       &                       &                      & aquatic & 0.66 & 0.84 & 0.74 \\
\midrule
\multirow{3}{*}{15} & \multirow{3}{*}{1024} & \multirow{3}{*}{2560} & \multirow{3}{*}{66.2} & host & 0.77 & 0.80 & 0.78 \\
                    &                       &                       &                      & soil & 0.49 & 0.45 & 0.47 \\
                    &                       &                       &                      & aquatic & 0.65 & 0.67 & 0.66 \\
\midrule
\multirow{3}{*}{10} & \multirow{3}{*}{2048} & \multirow{3}{*}{2560} & \multirow{3}{*}{65.5} & host & 0.77 & 0.81 & 0.79 \\
                    &                       &                       &                      & soil & 0.51 & 0.54 & 0.52 \\
                    &                       &                       &                      & aquatic & 0.67 & 0.61 & 0.64 \\
\midrule
\multirow{3}{*}{15} & \multirow{3}{*}{2048} & \multirow{3}{*}{1280} & \multirow{3}{*}{64.8} & host & 0.79 & 0.78 & 0.78 \\
                    &                       &                       &                      & soil & 0.50 & 0.53 & 0.51 \\
                    &                       &                       &                      & aquatic & 0.64 & 0.62 & 0.63 \\
\bottomrule
\end{tabularx} 
\end{table*}

\section{Overview of existing methods}\label{app:modelcomparison}

We compiled \cref{tab:comparison} summarizing the capabilities of most of the other potentially competing existing methods that we described in the related work. We assess them along the key requirements we set for our method, namely a) whether functional or taxonomic annotations/matches in existing databases are needed (often scarce for microbial life), b) whether they make use of full (coding) sequence information and scale to the full (coding) genome, and c) whether they allow for gene (interaction) attribution (requiring some sort of assessment of the influence or importance of all possible gene pairs on the prediction). 

In essence, existing approaches primarily fall short in at least one of the following two ways: a) They do not take into account the full sequence information but only highly abstracted annotations of individual genes. In this work, we consider all coding regions of a genome to qualify as ``full sequence'' as well.
These approaches typically do not have enough information for strong prediction performance, especially for prokaryotes where much less is known about a much larger fraction of organisms, c.f.\ ``microbial dark matter''.
b) They do not allow for pair-wise attribution, either because they have an excessively fine-grained granularity \citep{nguyen2023hyenadna,rojascarulla2019genet}, which makes gene-level identification intractable, or they accommodate large sequences via ``incomplete'' attention computations \citep{zaheer2021big,beltagy2020longformer}.
Most approaches to increase the maximally allowed input sequence length of transformers is by reducing the attention computation such that the quadratic cost is reduced (typically) to scale roughly linearly in the sequence length.
This inevitably means that we do not get all pairwise attention scores within any given sample, which is what our attribution method is built on.
If we relied on one of the linear attention methods, we would have to work with approximate/incomplete attributions as well---leading to missing relevant interactions only present in a subset of examples.
Another way of reducing the computational cost is by compressing the input sequences in the first place, e.g., via strided convolutions or other techniques to compress sequences \citep{Avsec2021-xz,benegas2023dna,Linder2023.08.30.555582}.
Instead of concatenating all gene sequences and compressing them jointly (thereby typically losing information about gene boundaries), our approach leverages existing large protein models to preserve genes as individual entities (but in a fixed-size vector representation instead of the base pair sequence).

\newcommand{\yes}{{\raisebox{0pt}[0pt][0pt]{\ding{51}}}}
\newcommand{\no}{{\raisebox{0pt}[0pt][0pt]{\ding{55}}}}
\newcommand{\maybe}{{\raisebox{0pt}[0pt][0pt]{(\ding{51})}}}

\begin{table*}
\centering 
\caption{Comparison of existing models in the literature with respect to the relevant aspects in our problem setting. The ``partial'' symbol {(\ding{51})} refers to the following. \emph{using full sequence}: HyenaDNA up to 1m bps; Borzoi up to 524k bps. \emph{attribution}: only for fragments of genes that cannot be pre-selected; \emph{classification}: adaptations to the original model are required; \emph{Using full sequence} means working with sequence level information directly and includes both the full genome as well as all coding regions.}\label{tab:comparison} 
\begin{tabularx}{0.9\textwidth}{c *{3}{C} c}
\toprule
\multicolumn{1}{c}{\textbf{model}} &
\multicolumn{1}{C}{\textbf{annotations required}} &
\multicolumn{1}{C}{\textbf{using full sequence}} &  
\multicolumn{1}{C}{\textbf{gene attribution}} & 
\multicolumn{1}{c}{\textbf{reference}} \\
\midrule
ours                & \no  & \yes   & \yes    & -- \\ 
baselines           & \no  & \yes   & \no     & described in \cref{sec:results}\\ 
HyenaDNA            & \no  & \maybe & \maybe  & \citep{nguyen2023hyenadna} \\
Enformer            & \no  & \no    & \maybe  & \citep{Avsec2021-xz} \\
Genomic Interpreter & \no  & \no    & \maybe  & \citep{li2023genomic} \\
Kraken 2            & \yes & \yes   & \no     & \citep{wood2019improved} \\
Traitar             & \yes & \no    & \yes    & \citep{weimann2016genomes} \\
BacPaCS             & \yes & \no    & \yes    & \citep{10.1093/bioinformatics/bty928} \\
Genet               & \no  & \no    & \maybe  & \citep{rojascarulla2019genet} \\
DNABERT             & \no  & \no    & \no     & \citep{Ji2020.09.17.301879} \\
Geneformer          & \no  & \no    & \no     & \citep{Theodoris2023-ok} \\
Borzoi              & \no  & \maybe & \maybe  & \citep{Linder2023.08.30.555582} \\
\bottomrule
\end{tabularx} 
{\footnotesize }
\end{table*} 

\section{Cluster ortholog annotations}\label{app:orthologs}

In \cref{tab:orth_host,tab:orth_soil,tab:orth_aquatic} we list all found orthologs from all the clusters in the three different habitats shown in \cref{fig:umaps}.
We provide this list as it demonstrates how our method can produce compact results that can be used by domain experts to inform their experiments and provide hypotheses for relevant interactions.
For concrete instances, one can swiftly look up these orthologs in databases (with usable online tools available) to get an idea of which genes have been clustered and which are important hubs within our gene interaction networks.

\begin{table}
\centering 
  \caption{Host cluster gene orthologs.}\label{tab:orth_host} 
\begin{tabularx}{\columnwidth}{lX}
\toprule
\multicolumn{1}{c}{\textbf{cluster}} &
\multicolumn{1}{c}{\textbf{KEGG orthologs}} 
\\ 
\midrule
blue & \texttt{K01992, K11051, K01095, K02950, K02887, K03628, K02992, K02952, K03438, K02986, K02874, K02358, K03686, K06168, K02913, K02988, K06217, K04077, K01338, K03544} \\
\midrule
orange & \texttt{K01537, K03043, K01624, K02945, K03553, K00611, K03496, K00088, K02976, K14623, K07254, K00549, K18929, K03621} \\
\midrule
green & \texttt{K02913, K02945, K03665, K06958} \\
\midrule
red & \texttt{K02935, K00817, K02950}\\
\midrule
purple & \texttt{K06898, K01937, K01677, K04565, K03628, K01939, K00052 ,K07246, K00097, K22024, K06334, K00937, K02996, K03816, K00533, K03070, K02431, K18843, K01571, K09124, K01892, K00335, K03658, K03086, K00773, K00640} \\
\midrule
brown & \texttt{K04751, K04752, K03628, K00573}\\
\midrule
pink & \texttt{K02886, K07448, K03106}\\
\midrule
grey & \texttt{K06996, K03856}\\
\bottomrule
\end{tabularx} 
\end{table}

\begin{table}
\centering 
  \caption{Aquatic cluster gene orthologs.}\label{tab:orth_aquatic} 
\begin{tabularx}{\columnwidth}{lX}
\toprule
\multicolumn{1}{c}{\textbf{cluster}} &
\multicolumn{1}{c}{\textbf{KEGG orthologs}} 
\\ 
\midrule
blue & \texttt{K08226, K02200, K07712, K01578, K00937, K10716, K06916, K17226, K00567, K00873, K07304, K07313, K01947, K03525, K02045, K11712, K10943, K01104, K02844, K14335, K03628, K06929, K03684, K00570, K03753, K04096, K01430, K01939, K08483, K09984, K01259, K03825, K07068, K10912, K08311, K03806, K08929, K05982, K18092, K17227, K01772, K00077, K02498, K00052, K02313, K08963, K07636, K00097, K22024, K00147, K01972, K07667, K09888, K01525}
 \\
\midrule
orange & \texttt{K01537, K03644, K03665, K00937, K01533K17686, K06916, K12297, K01626, K03695, K02017, K10763, K01578, K00254, K00931, K01534, K00574, K00773, K00325, K06921, K06954, K03723, K03466, K03786, K03655, K03656, K03694, K03555, K02669, K14682, K13821}\\
\midrule
green & \texttt{K00548,K01533, K17686, K01534, K01649} \\
\midrule
red & \texttt{no annotation}\\
\midrule
purple & \texttt{K03321, K14518, K02711, K00627, K00645, K01572, K02160, K09966} \\
\midrule
brown & \texttt{K01535, K01537, K01537}\\
\midrule
pink & \texttt{K15012, K02112, K01561, K01626, K06861, K02010, K02017, K00554, K03644, K06217, K00937, K01996} \\
\midrule
grey & \texttt{no annotation}\\
\bottomrule
\end{tabularx} 
\end{table}

\begin{table}
\centering 
  \caption{Soil cluster gene orthologs.}\label{tab:orth_soil} 
\begin{tabularx}{\columnwidth}{lX}
\toprule
\multicolumn{1}{c}{\textbf{cluster}} &
\multicolumn{1}{c}{\textbf{KEGG orthologs}} 
\\ 
\midrule
blue & \texttt{K01535, K01531, K17686, K01533, K17686, K00937, K03644, K00567, K22319, K00873, K16329, K07568, K14415, K03657, K03750, K07219, K13599, K07146, K02428, K03495, K12132, K11212, K00574, K08256, K00226, K00254, K01006, K01921, K01588, K15371, K06442, K00641, K07020, K14414, K03183, K01939, K07646, K01812, K01835, K01840, K07566, K14652, K00260, K00261, K01972, K00471, K00955, K05838, K06949, K00794, K14941, K01903, K03526, K07738, K00548, K01338}
 \\
\midrule
orange & \texttt{K21020, K01768, K07712, K07713, K07588, K02584, K06714, K07659, K05962}
\\
\midrule
green & \texttt{K04750, K01246} \\
\bottomrule
\end{tabularx} 
\end{table}

\section{Gene interaction networks}\label{app:gin}

\begin{figure}
    \centering
    \includegraphics[width=0.8\columnwidth]{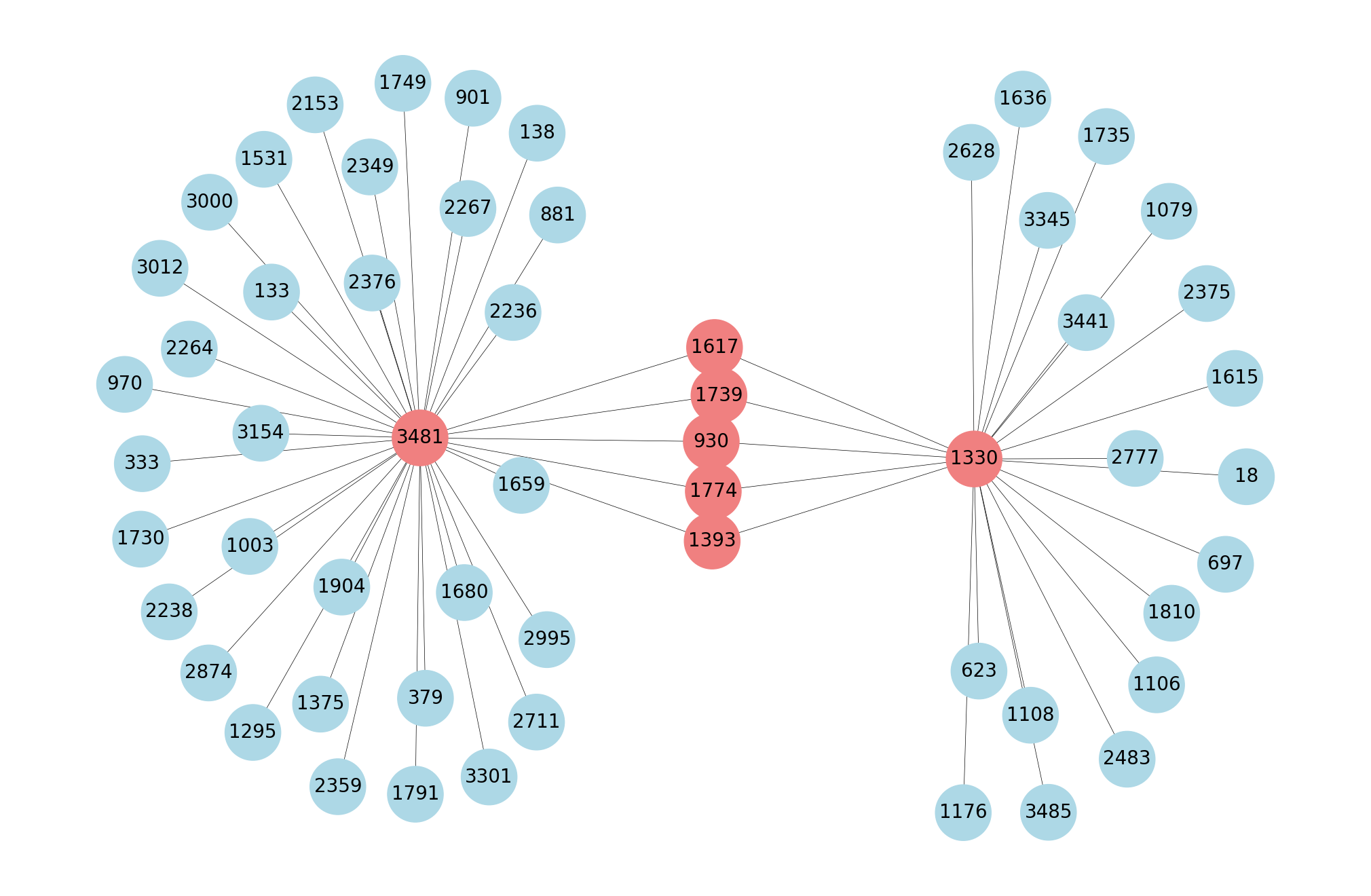}
    \caption{Gene interaction network constructed for the sample 159733.SAMEA6070310. Coral color indicates genes with more than one neighbor (hub), while blue indicates genes with only one connection (peripheral). Genes are numbered by the order of their appearance on the genome.}
    \label{fig:pathway2}
\end{figure}

\begin{figure}
    \centering
    \includegraphics[width=0.8\columnwidth]{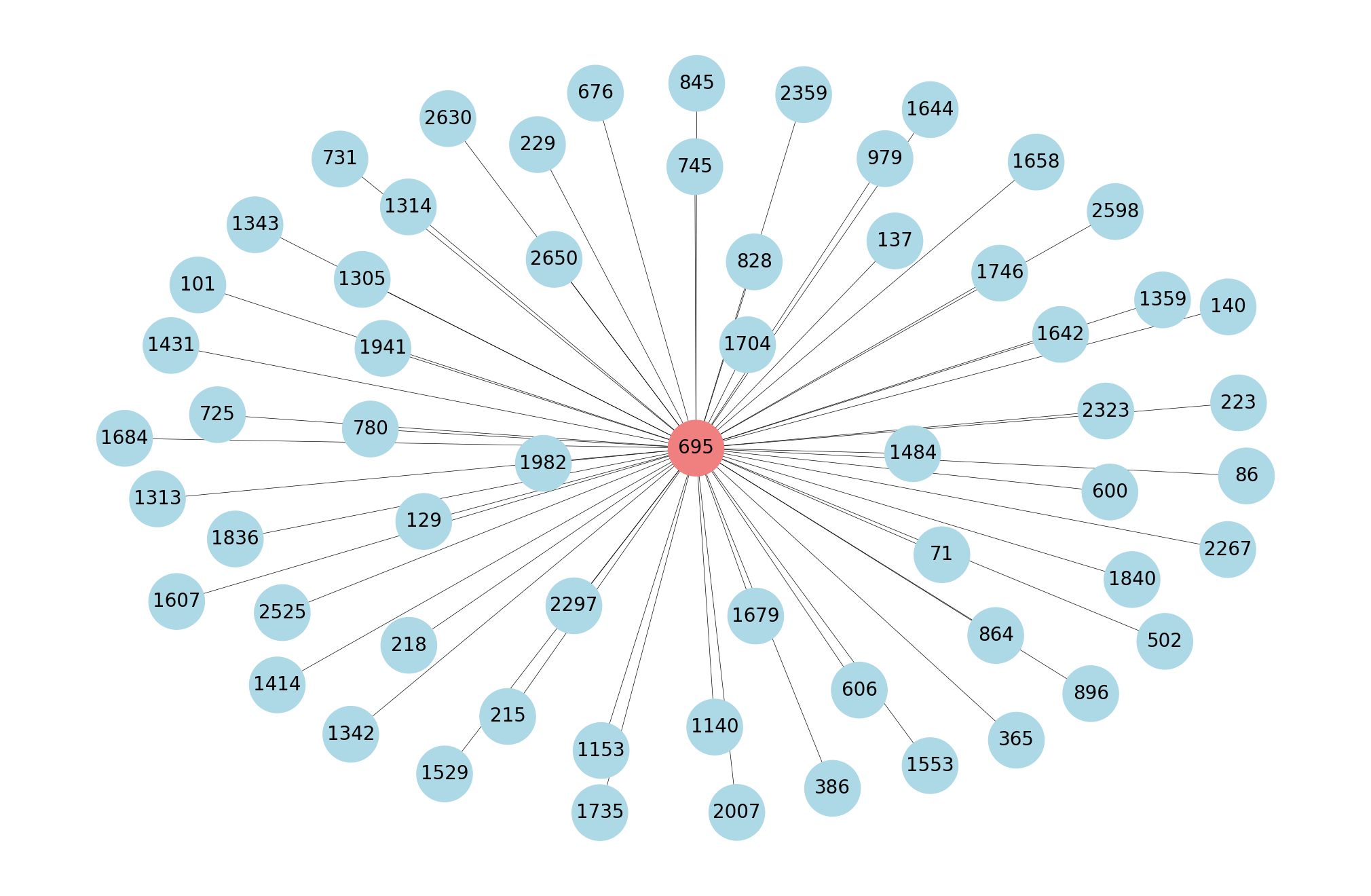}
    \caption{Gene interaction network constructed for the sample 2024894.SAMN08179843. Coral color indicates genes with more than one neighbor (hub), while blue indicates genes with only one connection (peripheral). Genes are numbered by the order of their appearance on the genome.}
    \label{fig:pathway3}
\end{figure}

We provide two additional examples of gene interaction networks from the aquatic and soil habitats. The network in \cref{fig:pathway2} is from an aquatic genome sample of Prochlorococcus marinus. The network in \cref{fig:pathway3} is from a soil genome sample of class Acidimicrobiia (unknown species).
Comparably, little can be said about the precise meaning and function of the key hubs in these networks, which highlights the fact that much less is known about free-living bacteria compared to the ones living in a host as they are comparably more relevant for human health and disease.
We thus leave these as two examples of relatively understudied and potentially interesting hypotheses to be followed up on experimentally.

\section{Impact statement}\label{app:impact}

This work presents methodological advances in the use of machine learning for predicting complex phenotypes from microbial genomic data, with potentially far-reaching implications for both the field of computational biology and society at large. By enabling more accurate predictions of habitat specificity from the genetic makeup of microbiomes and especially understanding the underlying drivers in terms of gene interactions, our research may aid innovative applications in environmental conservation, sustainable agriculture, and personalized medicine. The ability to understand and predict the interactions between microbial genes and their environments could lead to breakthroughs in the development of new biomarkers for health conditions, the creation of targeted microbiome therapies, and the enhancement of biodiversity conservation strategies.

Ethically, while the potential for positive impact is vast, we recognize the importance of considering potential downsides, especially in applications related to human and planetary health.
The same understanding that may be leveraged for improved treatments, may also be used to discover or engineer particularly resistant pathological organisms.
Generally, the adoption of advanced machine learning techniques in genomics must be accompanied by efforts to prevent misuse and ensure equitable access to the benefits they bring.
We advocate for a continued democratic dialogue to address these challenges and ensure that the advancements in computational biology contribute positively to society and the environment.

\section{Resources}\label{app:code}

Our project heavily relies on available open source software packages and data sources which we list in \cref{tab:software}.
\begin{table*}
\centering
\caption{Overview of resources used in our work.}\label{tab:software}
\begin{tabularx}{0.9\textwidth}{lYl}
  \toprule
    \textbf{Name} & \textbf{Reference} & \textbf{License}  \\ \midrule
    ProGenomes v3 & \citep{fullam2023progenomes3} & [none found] \\
    Python          &  \citep{vanrossum2009python}
                    & PSF License \\
    PyTorch         &  \citep{paszke2019pytorch}
                    & BSD-style license \\
    Numpy           &  \citep{harris2020numpy}
                    & BSD-style license \\
    Pandas          &  \citep{reback2020pandas,mckinney-proc-scipy-2010}
                    & BSD-style license \\
    Jupyter         &  \citep{kluyver2016jupyter}
                    & BSD-style license \\
    Matplotlib      &  \citep{hunter2007matplotlib}
                    & modified PSF (BSD compatible) \\
    Scikit-learn    &  \citep{pedregosa2011scikit}
                    & BSD 3-Clause \\
    SciPy           &  \citep{virtanen2020scipy}
                    & BSD 3-Clause \\
    HuggingFace     &   \citep{wolf2020transformers}
                    & Apache 2.0 (BERT models) \\
    ESM             & \citep{lin2023evolutionary}
                    & MIT license \\
    SLURM           & \citep{yoo2003slurm}
                    & modified GNU GPL v2 \\
    Biopython       &   \citep{cock2009biopython}
                    & modified BSD 3-Clause \\
    networkx        & \citep{networkx}
                    & BSD 3-Clause \\
    umap            & \citep{mcinnes2018umap}
                    & BSD 3-Clause \\
    PyFasta         & -- 
                    & MIT license \\
    Prodigal        & \citep{Hyatt2010}
                    & GNU GPL v3.0 \\
    EggNog-mapper   & \citep{steinegger2017mmseqs2,buchfink2021sensitive,eddy2011accelerated,cantalapiedra2021eggnog,huerta2019eggnog}
                    & GNU AGPL v3.0 \\
    DeepSpeed       & \url{deepspeed.ai}
                    & Apache 2.0 License\\
    \bottomrule
\end{tabularx}
\end{table*}

{\small
\bibliography{references}
}

\end{document}